**ECTIL: Label-efficient Computational Tumour Infiltrating Lymphocyte (TIL) assessment in breast cancer: Multicentre validation in 2,340 patients with breast cancer**

Yoni Schirris*[1,9,13,+], Rosie Voorthuis*[2], Mark Opdam[2], Marte Liefaard[2], Gabe S Sonke[3], Gwen Dackus[2], Vincent de jong[2], Yuwei Wang[2], Annelot Van Rossum[2], Tessa G Steenbruggen[3], Lars C Steggink[4], Liesbeth G.E. de Vries[5], Marc van de Vijver[6], Roberto Salgado[7,8], Efstratios Gavves[1], Paul J van Diest[11], Sabine C Linn[2,3,11], Jonas Teuwen[1,9], Renee Menezes[10], Marleen Kok[3,12] and Hugo Horlings[13]

*Joint first authors

1 University of Amsterdam, Science Park 402, 1098 XH Amsterdam, The Netherlands
2 Division of Molecular Pathology, Netherlands Cancer Institute, Amsterdam, The Netherlands
3 Department of Medical Oncology, Netherlands Cancer Institute, Amsterdam, The Netherlands
4 Department of Medical Oncology, Erasmus University Medical Center, Rotterdam, The Netherlands
5 Department of Medical Oncology, University Medical Center Groningen, The Netherlands
6 Department of Pathology, The Amsterdam University Medical Centre, AUMC, The University of Amsterdam, The Netherlands
7 Department of Pathology, ZAS Hospitals, Antwerp, Belgium
8 Division of Research, Peter Mac Callum Centre, Melbourne, Australia
9 AI for Oncology, Netherlands Cancer Institute, Plesmanlaan 121, 1066 CX Amsterdam, The Netherlands
10 Biostatistics Centre, Department of Psychosocial Research and Epidemiology, Netherlands Cancer Institute, Amsterdam, The Netherlands
11 Division of Pathology, University Medical Center Utrecht, Utrecht, The Netherlands
12 Department of Tumour Biology and Immunology, Netherlands Cancer Institute, Amsterdam, The Netherlands
13 Department of Pathology, Netherlands Cancer Institute, Amsterdam, The Netherlands

+ Corresponding author. h.horlings@nki.nl. +31-20-512-9111.

**Authors' contributions**

Yoni Schirris - conceptualisation, data curation, formal analysis, methodology, software, visualisation, writing - original draft, writing - review & editing, validation, investigation | Rosie Voorthuis - conceptualisation, data curation, resources, writing - original draft, writing - review & editing, investigation, visualization | Mark Opdam - resources, data curation, writing - review & editing | Marte Liefaard - resources, writing - review & editing | Gabe S Sonke - resources, writing - review & editing | Gwen Dackus - resources, writing - review & editing | Vincent de jong - resources, writing - review & editing | Yuwei Wang - resources, writing - review & editing | Annelot Van Rossum - resources, writing - review & editing | Tessa G Steenbruggen - resources, writing - review & editing | Lars C Steggink - resources, writing - review & editing | Liesbeth G.E. de Vries - resources, writing - review & editing | Marc van de Vijver - resources, writing - review & editing | Roberto Salgado, writing - review & editing, data curation | Efstratios Gavves, writing - review & editing, supervision, funding acquisition | Paul J van Diest - resources, writing - review & editing | Sabine Linn - resources, writing - review & editing, supervision | Jonas Teuwen - resources, software, writing - review & editing, supervision, funding acquisition | Renee Menezes - writing - review & editing, software, formal analysis, resources, supervision | Marleen Kok - conceptualisation, writing - review & editing, supervision | Hugo Horlings - conceptualisation, funding acquisition, data curation, writing - original draft, writing - review & editing, supervision, validation, project administration

**Abstract**

Introduction

The level of tumour-infiltrating lymphocytes (TILs) is a prognostic factor for patients with breast cancer and reflects their immune response, particularly in triple-negative breast cancer (TNBC). Computational TIL assessment (CTA) has the potential to assist pathologists in this labour-intensive task, since it can be quick and reproducible. However, current CTA models still rely heavily on detailed annotations, and use complex deep learning pipelines which pose challenges for model iterations and clinical deployment. Here, we propose and validate a fundamentally simpler deep learning based CTA that is trained in only ten minutes on hundredfold fewer pathologist annotations.

Methods

We collected whole slide images (WSIs) with TILs scores and clinical data of n=2,340 patients with breast cancer, including n=902 TNBC, from six cohorts including three randomised clinical trials. Morphological features were extracted from whole slide images (WSIs) using a pathology foundation model. Our label-efficient Computational stromal TIL assessment model (ECTIL) directly regresses the WSI TILs score from these features.

Findings

ECTIL trained on only a few hundred samples (ECTIL-TCGA) showed concordance with the pathologist over five heterogeneous external cohorts (r=0·54-0·74, AUROC=0·80-0·94). Training on TNBC slides only from four cohorts (ECTIL-TNBC) or all slides of five cohorts (ECTIL-combined) improved results, with ECTIL-combined achieving the highest concordance on an external test set (r=0·69, AUROC=0·85). Multivariable Cox regression analyses indicated that every 10% increase of ECTIL-combined TILs scores was associated with improved overall survival (OS) independent of clinicopathological variables (hazard ratio, 0·86 [0·78-0·94], p<0·01), similar to the pathologist score (hazard ratio 0·87 [0·82-0·92, p<0·001).

Interpretation

In conclusion, our study demonstrates that our proposed label-efficient ECTIL scores TILs on H&E FFPE WSI in a single step shows high concordance with an expert pathologist. Without using deep learning based segmentation and detection pipelines, ECTIL reached similar hazard ratios in an OS analysis independent of clinicopathological variables. In the future, such a CTA may be used to pre-screen patients for, e.g., prospective de-escalation trials or immunotherapy clinical trial inclusion or can be used as a tool to assist pathologists and clinicians in the diagnostic work-up of patients with breast cancer. Furthermore, our model is available online under an open source licence, allowing translational researchers to validate and use ECTIL in future studies in breast or other cancers.

Funding

Research at the Netherlands Cancer Institute is supported by institutional grants of the Dutch Cancer Society and the Dutch Ministry of Health, Welfare and Sport. The collaboration project is co-funded by the PPP Allowance made available by Health Holland, Top Sector Life Sciences & Health, to stimulate public-private partnerships.

## Research in context

### Evidence before this study

We searched Google Scholar and PubMed on 01 November 2023 for the search terms *("Deep Learning" OR "Artificial Intelligence") AND "Pathology" AND ("lymphocytes" OR "stromal tumour infiltrating lymphocytes" OR "TILs") AND "breast cancer"*, and reviewed the first 30 results. We included papers that validated deep learning applications to quantify stromal tumour infiltrating lymphocytes (TILs) in invasive breast cancer samples from haematoxylin and eosin (H&E) whole slide images (WSIs) (n = 11), and reviews thereof (n=1). The reviewed methods quantify the TILs to match pathologist scores or design novel lymphocyte-based metrics to optimise for prognostic or predictive performance. The published methods are concordant with pathologists (Spearman correlations ranging from 0·57 to 0·79 over studies and cohorts) and proved to have prognostic value independent of other clinicopathological variables. At the same time, their prognostic performance remains lower than that of a pathologist. All externally validated methods are pipelines of multiple steps that perform segmentation and detection to combine pixel-level information into the final biomarker, trained on very large numbers of costly tissue-level and cell-level annotations by expert pathologists.

### Added value of this study

This is the first study that externally validates a model with a fundamentally different design without detection and segmentation deep learning pipelines. Our model is trained solely on WSIs level TILs scores, in contrast to models trained using pixel-level semantic segmentations and cell annotations. Our model requires only a few hundred pathologist annotations to be trained; a reduction by two orders of magnitude compared to published methods using tens of thousands of annotations. Despite its few training labels, simplicity, and lack of any deep-learning-based segmentation mask or cell detections, the model reaches similar concordance (as measured by Spearman correlations ranging from 0·47-0·72 across other cohorts) and prognostic value independent of other clinicopathological variables in a cohort of node-negative, (neo)adjuvant chemotherapy-naive patients <40 years of age with TNBC. This is the largest retrospective validation study (n=2,340 over five cohorts) of any computational TILs assessment method. Our model performs well across various demographics, morphological and molecular subtypes, and medical centres, in both biopsies and resections.

### Implications of all the available evidence

This paper shows that our simple model is very data-efficient while showing high concordance with the pathologist, as well as prognostic value independent of other known prognostic clinicopathological variables. Both concordance and prognostic value are similar to existing methods that utilise multi-step segmentation-detection pipelines, while our model reduces annotation requirements by hundredfold. This provides evidence that data-hungry detection and segmentation pipelines may not be necessary for computational TIL assessment. Some, but not all, failure cases of existing methods are solved by our model. But compared to other complex, multi-step, detection segmentation pipelines, solving these known failure cases only requires a few additional labels and simple architectural changes. Our model will be prospectively validated in a de-escalation phase III clinical trial in patients with low-stage TNBC.

## Introduction

Breast cancer is the second-leading cause of cancer-related mortality in women worldwide.[1] Research on tumour immune interactions has identified tumour infiltrating lymphocytes (TILs) as a key prognostic and predictive biomarker across all breast cancer subtypes.[2] In triple-negative breast cancer (TNBC), TIL levels correlate with pathological complete response, disease-free-, and overall survival (OS).[2–4] Moreover, high TIL levels (>50% in patients >40 years of age and >=75% in patients <40 years of age) at baseline may indicate an excellent OS in stage 1 TNBC without (neo) adjuvant chemotherapy.[5,6]

TILs are determined through visual assessment of pathologists by dividing the ratio of TILs by the total stromal tumour area on hematoxylin- and eosin-stained (H&E), formalin-fixed paraffin-embedded (FFPE) whole-slide images (WSIs) following the guideline of the International Immuno-Oncology Biomarker Working Group on Breast Cancer (TILs-WG).[7] While the scoring process and additional training have enhanced reproducibility among pathologists, inter- and intra-observer variability persists as a challenge , partly due to intra-tumour heterogeneity, as shown by Kos et al. with intraclass correlation coefficients ranging from 0·77-0·94 over three study groups.[8]

Recent advancements in Artificial Intelligence (AI) and deep learning (DL), coupled with digitalisation to WSIs, have driven the development of Computational TILs assessment (CTA) methods. Furthermore, the TIL-WG has introduced computational guidelines to address the full complexity of the scoring process, aiding the development of useful CTA that may be adopted for clinical use.[9] Moreover, AI-assisted TILs assessment could potentially offer some solutions to the challenges faced by pathologists, including diminishing the burden on pathologists and pathology departments.

Current models typically offer explicit pixel-level segmentations and cell detection and classifications.[10–13] While these models demonstrate high concordance with the pathologists' TILs score and are associated with improved survival, independently of other known prognostic factors, they still face numerous pitfalls and challenges as described by Thagaard et al.[14] Existing CTAs generally encounter difficulties due to various technical slide issues, and adhering to the pathologist scoring guidelines remains challenging.[9] Moreover, they demand significant amounts of labelled data for training, which is costly and time-consuming. As these models become more complex to provide pixel-level information, continuous validation of all submodules in clinical practice may become increasingly difficult. Despite providing pixel-level segmentation and cell locations while explicitly implementing scoring guidelines, these models still encounter discordant cases due to, e.g., inclusion of incorrect tissue areas or cells.

In this study, we present an AI-based CTA that requires two orders of magnitude (hundredfold) fewer annotations than existing methods. By utilising only a few hundred WSI-level TILs scores as training data, we develop a model that regresses the TILs score in a single step from the WSI without post-processing, and we validate this model in the largest CTA validation study to date. We extend an existing DL-based WSI-level TILs regression method with an improved foundational model, gated attention mechanism, and additional feature learning.[15] ECTIL (Efficient Computational stromal TIL assessment, pronounced "Easy-TIL") is a straightforward, easy-to-train tool for accurately identifying and quantifying TILs on H&E-stained WSIs of either biopsies or full resections of patients with breast cancer across molecular and morphological subtypes. The model directly regresses the TILs score from the WSI without explicit implementation of the TILs-WG scoring guideline. Even when ECTIL is trained on only a few hundred samples from The Cancer Genome Atlas Breast Cancer cohort (TCGA), which finishes training in only 10 minutes on a commodity graphics processing unit, ECTIL demonstrates concordance with pathologists' TILs scores over five external cohorts, indicating strong generalisation capabilities. Additionally, the ECTIL score provides prognostic value independent of clinicopathological variables on a single external cohort. ECTIL significantly reduces the label requirements by more than hundredfold compared to models trained on, e.g., the TIGER grand challenge, which utilise 30,000 lymphocyte and plasma cell annotations and 300 densely annotated regions with pixel-level tissue information. ECTIL has the potential to streamline and automate the TILs scoring process in breast cancer, facilitate the use of TILs scores as a biomarker in translational research and ultimately enhance its integration into clinical practice. Furthermore, ECTIL will be integrated in a prospective de-escalation phase III clinical trial in patients with low stage TNBC.

## Methods

### Study design and cohorts

Individual clinical and histopathological data including WSI from primary tumour material with an available pathologist TILs score was collected from six cohorts of patients with early breast cancer. Our ethical considerations are stated in supplementary section S1. Two cohorts consist of publicly available data. The Cancer Genome Atlas (TCGA) is a large cancer genomics dataset that has molecularly characterised primary cancer samples of different cancer types, including breast cancer (n=1,134). The clinical variables were exported from the Genomic Data Commons repository. The second open source cohort consists of data acquired through the Breast Cancer Somatic Genetics Study (BASIS, n=560) as published by Zainal et al.[16] Three cohorts consist of patients treated in randomised clinical trials (N4+ n=855, MATADOR n=664, TRAIN II n=438). As WSI were only included when a TILs score by an expert pathologist was already available, some patients were not included in this current study. The inclusion flow diagram is depicted in supplementary figure S1. A detailed description of each trial design is stated in supplementary section S2.

The PARADIGM cohort, which was used as external test set only, consists of clinical data and primary tumour material collected retrospectively from 481 young (<40 years), any T-stage, node-negative (N0), metastasis free (M0), systemic therapy naive patients with TNBC treated between 1989 and 2000 in The Netherlands collected as part of the PARADIGM initiative.[17]

### Clinicopathological characteristics

Clinicopathological data was received directly from the principal investigator of the study. Tumour grade (according to the Bloom-Richardson classification)[18], presence of lymphovascular invasion, breast cancer histology and subtype (hormone and HER2 receptor status) were assessed locally according to national guidelines,[19] or centrally, according to the study protocol.

### TILs

TILs were scored on H&E FFPE WSIs according to previously published recommendations by the TILs-WG.[7,20] All TILs scores were collected retrospectively. If an independent TILs score was available from two or more pathologists for one slide, the mean of the scores was used for training, validation, and analysis. There are cases where two WSIs and only one TILs score was available for a patient; in TCGA, both slides were included with the same label while for all other cohorts the WSI with the highest quality was used. TILs were scored on primary tumours that were removed during surgery prior to any neoadjuvant systemic therapy, except for the TRAIN II cohort where TILs were scored on the diagnostic biopsy.

### Patient and WSI selection workflow

WSI were collected from the studies in which a TILs score by the pathologist was already available, and there was compatible clinical data which in some cohorts reduced the number of patients that could be included in this current study (supplementary figure S1). No further patient or WSI selection was made. WSI were scanned by different scanners (Aperio, Hamamatsu C9600-12 or 3DHistech P1000). For the TCGA cohort 14 patients had two representative WSI which were both included in the ECTIL models. The WSI selection workflow is depicted in the supplementary material (figure S1). In short, we collected 5799 WSI over six cohorts. Pathologist's TILs scores were provided for 2480 patients. Sufficient clinical variables were available for 2340 patients with 2354 slides to perform concordance analyses. For the survival analyses in PARADIGM, six patients had insufficient survival data. Hence, 390 patients were included in the concordance analysis, while 384 patients were used in the survival analysis.

### Deep Learning Model: ECTIL

All steps of ECTIL are visualised in figure 1, and described in detail below.

**Tissue selection**

Background is discarded using the Foreground Extraction from Structure Information (FESI) method as presented by Bug et al. (2015), as implemented in the Deep Learning Utilities for Pathology (DLUP, v0·3·22) library developed in-house in the Netherlands Cancer Institute.[21,22] We remove any tile with 0 foreground pixels. Qualitative results of samples from each dataset are shown in figure S5-S7.

**Feature extraction**

We use RetCCL, a convolutional neural network that is pre-trained using self-supervised learning on pathology slides by Wang et al. to extract a 2048-dimensional feature vector from 512x512px patches at 0·5 microns per pixel with no overlap after normalisation using imagenet statistics.[23] RetCCL was chosen due to its superior performance shown by Niehaus et al.[24]

**Model**

We use a gated attention-based multiple instance learning (ABMIL) model to regress the TILs score from the extracted features of WSI patches directly based on previous work.[15,25] We add a non-linear layer with a rectified linear unit layer after the extracted feature vectors to generate a new task-specific tile encoding and a sigmoid layer to map the model output to [0,1]. The hyperparameter search and final parameter set is described in supplementary section S3, and they are used to train all models.

**Model training and evaluation**

ECTIL-TCGA is trained using five-fold cross validation split by medical centre of the TCGA cohort (n=356 slides). The model is trained on three folds, validated during training on one fold, and internally tested on the remaining fold. Finally, the model is externally tested on all other cohorts. ECTIL-combined is trained on all cohorts except PARADIGM (n=1964 slides) with a five-fold leave-one-cohort-out data split leaving one cohort for validation during training in each fold, and is tested only on PARADIGM. ECTIL-TNBC is trained on only TNBC samples from the TCGA, BASIS, N4+ and MATADOR (n=400 slides) with a four-fold leave-one-cohort-out data split leaving one cohort for internal validation, and is tested only on PARADIGM. For external testing of each ECTIL model, we pick the model with the highest Pearson's r on the validation set of each fold, using the ensemble of models for the test set.

**Statistical analysis**

We perform concordance analyses by computing the mean squared error (MSE), Pearson's-, Spearman-, and concordance correlation coefficient in Python using torchmetrics.[26] To analyse the model's capacity to identify high vs low TILs we split the pathologist score into TILs-low ($< 30\%$) and TILs-high ($\geq 30\%$). Using these binary ground-truth labels and the model's predicted TILs score, we compute the area under the receiver operating characteristic curve (AUROC). In the supplementary materials we additionally provide the average precision (AP) and AUROC for varying cutoff points (10%, 50%, 75%). We visualise model predictions over datasets and slides using scatter plots, calibration curves, and score- and attention heatmaps.

We reproduce the OS analysis as reported in de Jong et al. in R using the *survival* package on the PARADIGM cohort.[5] To test the impact of the pathologist's and ECTIL TILs score on OS we calculate hazard ratios (HR) and concordance indexes (CI) using uni- and multivariable Cox proportional models, with adjustment for other clinicopathological characteristics (including T stage, tumour grade, lymphovascular invasion and local treatment) in the PARADIGM cohort. The ECTIL score is normalised to the minimum and maximum of the training dataset. We produce Kaplan Meier plots for the pathologist's score with previously defined cutoffs (30%, 75%).[5] To compare the survival split of the pathologist to ECTIL, we split both scores by their median. The Schoenfeld residuals are used to test the proportionality of hazards assumption over time.

## Results

### Study population

The baseline characteristics of all collected cohorts are set out in table 1, with a total of 2,340 patients with high quality WSI and complete clinical data. Since the PARADIGM cohort consists of a very specific patient population, and it is used for prognostic analyses, this cohort is described separately. The median age over the cohorts excluding PARADIGM (table S1, ECTIL-Combined column) is 49 (min-max = 22-90) years. The majority of patients are female (98·9%). Most patients had stage II (42·9%) or III (44·1%) disease at time of diagnosis or study entry. Ten patients (0·5%) had known stage IV disease at study entry. Most tumours were categorised as breast cancer of no special type (BC NST, 66·6%) with either tumour grade II (33·4%) or III (34·4%). Presence or absence of lymphovascular invasion was unknown for the majority of tumours (88·7%). Most tumours were ER+/HER2- (59·8%), 11·7% of the tumours were ER+/HER2+, 6·3% ER-/HER2+ and 20·4% were TNBC. The median pathologist TILs score was 10% (min-max: 0-92·5).

For the PARADIGM cohort, the median age was 35 years (min-max = 22-39). All patients were female and premenopausal. Stage I disease was most prevalent (59·5%), followed by stage II (40·3%). All patients were node negative at diagnosis, with no known distant metastasis. The majority of tumours were BC NST (85·4%). Twenty (5·1%) patients had metaplastic carcinoma. The majority of tumours were grade 3 (86·2%), followed by grade 2 tumours (13·3%). Lymphovascular invasion was present in 10·3% of the tumours. All tumours in PARADIGM were TNBC. The median TILs score was 20% (min-max:1%-95%).

An overview of the baseline characteristics of the ECTIL-TNBC and ECTIL-combined cohorts, which are used for training ECTIL-TNBC and ECTIL-combined models, are shown in table S1.

### ECTIL-TCGA is concordant with the pathologist in external cohorts

First, we compared the predictions of ECTIL-TCGA to the pathologist scores in the TCGA internal test set, BASIS, N4+, TRAIN II, MATADOR, and PARADIGM cohorts. When evaluated on the internal test set of the TCGA cohort, ECTIL-TCGA achieved a Pearson's r of 0·61 and AUROC of 0·84 (table S2). In figure 2 we see that ECTIL-TCGA achieves, across external test cohorts, a Pearson's r of 0·54-0·74 and an AUROC of 0·80-0·94. ECTIL achieved the highest concordance on MATADOR (r=0·74, AUROC=0·94), and the lowest on PARADIGM (r=0·58, AUROC=0·80) or N4+ (r=0·54, AUROC=0·82) depending on the metric of interest. Additional metrics, and the AUROC for varying pathologist score cutoffs can be found in table S2, and case-level predictions are presented through scatter plots in figure S2.

### Training ECTIL on multiple cohorts improves external validation performance on PARADIGM

Second, we evaluated ECTIL-TNBC and ECTIL-combined on the PARADIGM cohort. ECTIL-TNBC,, reaches a higher performance (r=0·64, AUROC=0·83) on the PARADIGM cohort compared to ECTIL-TCGA (r=0·58, AUROC=0·80) (figure 2). ECTIL-combined, which includes all slides from all cohorts except PARADIGM (n=1,964), reached a Pearson's r of 0·69 and an AUROC of 0·85 (figure 2). Additional metrics and the AUROC for varying pathologist score cutoffs can be found in table S3.

To further investigate the predictions of ECTIL-combined on the PARADIGM cohort we constructed a calibration plot in figure 3 (for ECTIL-TCGA in figure S3, and for ECTIL-TNBC in figure S4). We plotted the pathologist score distribution (y-axis) conditioned on the predicted scores of ECTIL-combined in bins of size 0·05 (x-axis). The distribution of ECTIL-combined scores was smaller (0-0·7), compared to the pathologist score (0·1-0·95). Moreover, ECTIL-combined tended to overpredict the samples with a low (< 0·2) pathologist score, and underpredicts samples with a higher (> 0·3) pathologist score. Additionally, it became clear that there are discordant cases. For each sample with a pathologist score of 0·9, ECTIL may predict these to have a value of 0·2-0·7. Samples that were predicted to have a value of 0·2-0·5 may contain a sample that is originally scored by a pathologist to have, e.g., a TILs score of 0 or 0·9.

**Discordant case analysis highlights model limitations**

Next, we qualitatively analysed samples that received highly different scores from ECTIL compared to the pathologist to understand ECTIL's limitations. For each cohort and for two groups (pathologist TILs score $< 30$ and $> 50$, which we term pTIL-l and pTIL-h respectively) we randomly choose one sample from the top ten samples with the largest differences between the predicted score and the pathologist score. For these samples, we visualise the heatmap of ECTIL-predicted scores and attention weights for each patch, which were analysed by a board-certified pathologist. These samples are presented in figure S5-S7-2. The first limitation was on samples from invasive lobular carcinoma where tumour cells were scored as TILs. Additionally, patches that contain lymphovascular invasion, other immune hotspots, fibrotic area, and lymphocytes at the tumour boundary received a high TILs score and attention score, while a pathologist would not include these lymphocytes. Cases with large WSI and small tumour beds may receive a different score due to the accumulated attention weight of the large number of patches with low attention. In the aggregated results of the subset analysis presented in table S4, we find that ECTIL-TCGA indeed performed quantitatively worse on patients with invasive lobular carcinoma (r=0·36, AUROC=0·64) across datasets compared to, e.g., breast cancer of no special type (BC NST) (r=0·58, AUROC=0·73).

**The ECTIL score is prognostic independent of other known prognostic clinicopathological variables**

Finally, we determined whether the predicted TILs scores by ECTIL-TCGA, ECTIL-TNBC and ECTIL-combined are independently associated with OS. In the univariable Cox regression analysis with TILs in 10% increments, all three ECTIL TILs scoring models yield statistically significant HR (0·87 [0·79-0·96], p<0·01 for ECTIL-TCGA, HR 0·86 [0·78-0·94], p<0·01 for ECTIL-combined, HR 0·85 [0·77-0·94], p<0·01 ECTIL-TNBC), similar to the pathologist TILs score (HR 0·87 [0·82-0·92], p<0·001), as outlined in table 2. In the multivariable Cox regression the HR of 10% TILs increments of all ECTIL models (HR 0·87 [0·79-0·95], p<0·01 for ECTIL-TCGA, HR 0·85 [0·77-0·93], p<0·001 for ECTIL-combined, HR 0·84 [0·76-0·93], p<0·001 for ECTIL-TNBC) remain significant, independent of all other clinical characteristics, similar to the pathologist score (HR 0·86 [0·81-0·92], p<0·001), outlined in table 2.

We generated Kaplan-Meier curves for the PARADIGM cohort using various scoring methods, with pathologist TILs score cutoffs previously used in the literature (TILs scores of <30, 30-75, ≥75), pathologist scores with median cutoffs (TILs score of ≥ 20 vs. <20), and ECTIL-combined scores split by the median (TILs score of ≥ 27 vs. <27) (Supplementary figure S8, S9, S10 respectively). The analysis revealed significant differences in OS between the groups (pathologist TILs score of ≥75%, 30-74% and <30% have a 100%, 88.9% and 80.6% 3-years OS respectively; pathologist TILs score of ≥20% (median), 94% versus 80% 3-years OS; ECTIL-combined TILs score ≥27 (median), 90% versus 83% 3-years OS), highlighting the clinical value of both ECTIL and pathologist scores when split by the median in patients with systemic treatment naive low stage, N0, TNBC < 40 years of age. The HRs for these groups further supported these findings, indicating the validity of these scoring methods in predicting OS (Figure S8-S10 and table 2).

**Discussion**

In this study, we present ECTIL, a label-efficient CTA that regresses the TILs score from extracted features of WSI patches directly. To our knowledge, this is the largest retrospective validation study of a CTA thus far, including 2,340 patients with available WSIs of their breast carcinoma, TILs scores, and relevant clinical data. We demonstrated that ECTIL, when trained on a limited dataset of only 200 WSIs with a single pathologist annotation (% of TILs score), can produce scores that are highly concordant with a pathologist's assessments across multiple external test cohorts. This concordance is reflected in performance metrics, including Pearson's correlation and AUROC values, which indicate strong agreement with pathologist evaluation. Additionally, ECTIL-generated TILs scores show a significant association with OS, independent of other clinicopathological variables, suggesting that ECTIL can provide prognostic value comparable to that of pathologists. Our findings also reveal that expanding the training dataset to include larger and more heterogeneous cohorts enhances ECTIL's concordance with pathologist assessments. This improvement underscores the importance of training on diverse data sets to maximise model performance. Conversely, restricting the training data to a narrower subset, such as only patients with TNBC, diminishes ECTIL's effectiveness, further highlighting the benefit of diverse training data over homogeneity.

In contrast to existing models that generally use a detection-segmentation pipeline involving multiple models and postprocessing steps, ECTIL is a single model trained on two orders of magnitude fewer annotations which finishes training from scratch in only 10 minutes on a commodity graphics processing unit (NVIDIA GeForce RTX 2080 Ti).[10–13] Despite this fundamentally simple design, ECTIL reaches a concordance with the pathologist that is highly comparable to existing methods. For example, Bai et al.[11] obtain a Spearman's r of 0·61 and 0·63 on two TNBC cohorts, whereas we reach a Spearman's r of 0·47 to 0·72 across five cohorts with varying molecular subtype distributions, and 0·61 for our TNBC cohort. When representing the TILs score as a fraction (from 0 to 1), Sun et al.[10] report a root mean squared error (RMSE) of 0·15 to 0·22 across three models and two TNBC datasets in their supplementary materials, and ECTIL obtains a RMSE of 0·14 to 0·28 over more heterogeneous cohorts. Thagaard et al.[14] present the results of commercial software with a Spearman correlation of 0·79, a much higher score than other published scores. It must be noted, however, that this model was developed using immunohistochemistry-guided annotations on slides that were picked as a held-out training set from the cohort collected at their institution. Hence the generalizability to external cohorts cannot be compared easily. Finally, Choi et al.[13] present the results of commercial software reaching a concordance correlation coefficient of 0·71 and 0·75 on two TNBC datasets, which is higher than most coefficients achieved by ECTIL. These impressive results can only be achieved with a model trained on a staggering amount of training data, however - 2,712 WSI from multiple cancer types with a total of 2,334 $mm^2$ dense tissue type annotations with 558,225 explicit cell annotations were utilised. This insight is particularly relevant given that many existing computational tumour analysis models have traditionally relied on more homogeneous training sets. Our study advocates for the inclusion of diverse data in training regimens to enhance model performance and predictive accuracy. These results suggest that ECTIL is able to generalise and perform well across different cohorts and may be more suitable for implementation into daily clinical practice. Although a perfect comparison is not feasible, only commercial CTA methods with extensive annotations appear to outperform ECTIL. Given that ECTIL is trained on substantially fewer annotations, it effectively addresses the main challenge highlighted by Thagaard et al, which is the costly collection of annotations across diverse data distributions to develop generalizable algorithms.

Despite not explicitly following TILs-WG guidelines, ECTIL learns to predict TILs scores through supervised training with guideline-based scores, suggesting that direct translation of these guidelines into computational frameworks may not be absolutely necessary. ECTIL's validation across numerous cohorts with diverse patient characteristics, technical preparations, scanners, and extraction strategies demonstrates its generalizability. Through qualitative analysis of highly discordant cases by a board-certified pathologist who is trained to score TILs, we identified the limitations of ECTIL. Interestingly, the patch-level TILs scores match the pathologist's idea of the number of stromal lymphocytes, and the attention heatmaps often show a clear focus on mostly relevant tissue. The majority of discordant cases are caused by erroneously included areas like lymphocytic hotspots, lymphocytes at the tumour border, lymphocytes in necrotic areas, and lymphocytes in lymphovascular invasion. These issues are exacerbated when the tumour area is small relative to the WSI, since there is more incorrect tissue to include. Interestingly, however, the TRAIN II cohort, which only includes biopsies that have a low tumour-to-tissue ratio, does not lead to a significantly lower performance than ECTIL achieves on resections. Another clear challenge is scoring invasive lobular carcinomas (ILC) - which is a difficult task even for pathologists since the tumour cells have a similar morphology to lymphocytes and are potentially scored as such. Interestingly, these pitfalls (ILC, erroneous tissue inclusion) are highly similar to those in previous works as described by Thagaard et al., while our model design is fundamentally different.[14] ECTIL correctly identifies pen markings and air bubbles, which receive low attention and were not the cause of discordance in the discordant cases we analysed. A limitation that is unique to the architecture of ECTIL, however, is the dilution of scores in large WSIs with a small tumour bed. Due to the cumulative effect of a large number of tiles with a low attention score, the scores of these tiles may still contribute 50% of the total weight. This dilution is the main reason our model predicts a maximum score of 70%, which may only be solved using tumour bed annotations or using a context-aware model architecture.

While promising, ECTIL requires further refinement. Since it only needs WSI-level labels, training datasets can be easily enriched with additional labelled data. For example, in our cohorts only 5% of samples are ILC, contributing to lower performance in this subtype. Additionally, employing more recent foundation models like Phikon by Filiot et al. can enhance representation of the underlying morphology.[27] One of ECTIL's architectural limitations is that it treats patches independently and lacks context-awareness. As we know that context matters, future improvements could involve using context-aware models, such as vision transformers, which effectively handle WSIs. This architecture captures spatial information and enables it to better learn to follow TILS-WG scoring guidelines.

Before a CTA like ECTIL can ever be implemented in the clinic, however, it is required to utilise a separate model to perform quality control since even an image of only a glass slide with no tissue would be given a score, similar to existing models. Additionally, it is the responsibility of a pathologist and clinician to interpret the model's score and decide on the clinical impact for the patient. Before implementing ECTIL as a direct pre-screening tool or tool to assist the pathologist, prospective validation studies are required. In the near future, ECTIL will be incorporated in a prospective de-escalation phase III clinical trial in patients with low-stage TNBC.

In conclusion, ECTIL scores TILs on H&E FFPE WSI in a single step without using DL-based segmentation and detection pipelines. ECTIL is data-efficient and scores TILs highly similar to an expert pathologist. It also reaches a similar hazard ratio in an overall survival analysis independent of clinicopathological variables. In the future, such a CTA may be used to pre-screen patients for, e.g., prospective de-escalation trials or immunotherapy clinical trial inclusion or to be used as a tool to assist pathologists and clinicians in clinical decision-making.

## Additional Material

### Acknowledgements

The authors would like to thank the patients and their families for participation in the clinical studies and allowing them to make use of their data for further breast cancer research. Furthermore, the authors thank Beheerstichting Borstkanker Onderzoek Groep (BOOG) as supporter and coordinator of the MATADOR study and sponsor of the Train II study, together with all participating hospitals for patient inclusion, data collection and providing the availability of biospecimen. In addition, the authors thank the Core Facility Molecular Pathology & Biobanking at the Netherlands Cancer Institute for digitising the whole slide images and wish to acknowledge the Research High Performance Computing (RHPC) facility of the Netherlands Cancer Institute (NKI).

The results published here are in whole or part based upon data generated by the TCGA Research Network: https://www.cancer.gov/tcga

### Data Sharing

All the code required to train and validate the models is available at https://github.com/nki-ai/ectil, including the scripts to reproduce the performance metrics and prognostic analyses.

The WSIs and clinical data of The Cancer Genome Atlas are publicly available at https://portal.gdc.cancer.gov/ and https://www.cbioportal.org/, respectively. The pathologist TIL scores of TCGA samples are available on github (https://github.com/nki-ai/ectil). Open access to the WSIs and clinical data of BASIS are available for request at https://daco.icgc-argo.org/, and the pathologist TIL scores are available upon reasonable request.

N4+:
The datasets used and analysed during the present study are not publicly available due to privacy-protection restrictions, but are available from prof. dr. S. C. Linn upon reasonable request and in the presence of a data transfer agreement.

MATADOR:
The clinical data including TIL scores are publically available at (doi: 10.17632/n24jkp487r.2). The WSIs are not publicly available but may be requested for use from prof. dr. S. C. Linn in the presence of a data transfer agreement.

TRAIN II:
The data collected for this study can be made available to others in de-identified form after all primary and secondary endpoints have been published and in the presence of a data transfer agreement. Requests for data sharing can be made to prof. dr. G. Sonke, including a proposal that must be approved by the trial's steering committee.

Paradigm:
The datasets used and analysed during the present study are not publicly available due to privacy-protection restrictions, but are available from prof. dr. S. C. Linn upon reasonable request and with permission of the Netherlands Cancer Registry, hosted by the Netherlands Comprehensive Cancer Center (IKNL).

### Funding

Research at the Netherlands Cancer Institute is supported by institutional grants of the Dutch Cancer Society and the Dutch Ministry of Health, Welfare and Sport.

The collaboration project is co-funded by the PPP Allowance made available by Health Holland, Top Sector Life Sciences & Health, to stimulate public-private partnerships.

**Role of the funding source**

The funding source has had no role in the writing of the manuscript, the decision to submit it for publication, data collection, analysis, or interpretation, or any other immediate aspect pertinent to the study. None of the co(authors) has been paid to write this article by a pharmaceutical company or other agency. The authors were not precluded from accessing data in the study, and they accept responsibility to submit for publication.

**Declaration of interests**

**Yoni Schirris**

No known COI

**Rosie Voorthuis**

No known COI

**Mark Opdam**

No known COI

**Marte Liefaard**

No known COI

**Gabe S Sonke**

No known COI

**Gwen Dackus**

No known COI

**Vincent de jong**

No known COI

**Yuwei Wang**

No known COI

**Annelot GJ Van Rossum**

No known COI

**Tessa G Steenbruggen**

Payment or honoraria for podcasts outside the work submitted: Gilead Sciences.

**Lars C Steggink**

No known COI

**E.G.E. de Vries**

E.G.E. de Vries reports institutional financial support for advisory boards/consultancy from NSABP, Daiichi Sankyo, and Crescendo Biologics and institutional financial support for clinical trials or contracted research grants from Amgen, Genentech, Roche, Bayer, Servier, Regeneron, and Crescendo Biologics, all outside the submitted work.

**Marc van de Vijver**

No known COI

**Roberto Salgado**

Consulting or Advisory Role: Roche, AstraZeneca, BMS, Daiichi Sankyo

Speakers' Bureau: Exact Sciences, AstraZeneca, Daiichi Sankyo

Research Funding: Merck, Puma Biotechnology, Roche

Travel, Accommodations, Expenses: Roche, Merck, AstraZeneca, Daiichi Sankyo

**Efstratios Gavves**

Shareholder of Ellogon.ai

**Paul J. van Diest**

Consulting or Advisory Role: Visiopharm, Paige, Sectra

Research Funding: Paige, Pfizer, Proscia

**Sabine C Linn**

Consulting or Advisory Role: Daiichi Sankyo (Inst), AstraZeneca (Inst), Cergentis (pro bono), IBM (Inst).

Institutional Research Funding: Genentech/Roche, AstraZeneca, Tesaro (now GSK), Merck, Immunomedics (now Gilead), Eurocept Pharmaceuticals, Agendia, Novartis.

Travel, Accommodations, Expenses: Daiichi Sankyo Europe GmbH (Inst)

**Jonas Teuwen**

Shareholder of Ellogon.ai

**Renee Menezes**

No known COI

**Marleen Kok**

Research funding: BMS, Roche, Astrazeneca

Advisory roles compensated to the institute: Astrazeneca, Daiichi Sankyo, Domain Therapeutics, Alderaan, BMS, MSD, Roche

Speakers fee compensated to the institute: Roche, BMS, Gilead

**Hugo Horlings**

No known COI

## Tables

## Table 1: Baseline characteristics of all clinical cohorts

| | TCGA (N=356**) | BASIS (N=247) | N4+ (N=552) | MATADOR (N=517) | TRAIN II (N=292) | PARADIGM (N=390) |
|---|---|---|---|---|---|---|
| **Age (years)** | | | | | | |
| Median [Min, Max] | 50·0 [26·0, 90·0] | 57·0 [28·0, 81·0] | 46·0 [26·0, 55·0] | 51·0 [26·0, 72·0] | 49·0 [22·0, 74·0] | 35·0 [22·0, 39·0] |
| Missing | 14 (3·9%)** | 0 (0%) | 0 (0%) | 0 (0%) | 0 (0%) | 0 (0%) |
| **Gender** | | | | | | |
| Female | 338 (94·9%) | 244 (98·8%) | 552 (100%) | 517 (100%) | 292 (100%) | 390 (100%) |
| Male | 4 (1·1%) | 3 (1.2%) | 0 (0%) | 0 (0%) | 0 (0%) | 0 (0%) |
| Missing | 14 (3·9%)** | 0 (0%) | 0 (0%) | 0 (0%) | 0 (0%) | 0 (0%) |
| **Menopausal status** | | | | | | |
| Premenopausal | 0 (0%) | 58 (23·5%) | 0 (0%) | 266 (51·5%) | 160 (54·8%) | 390 (100%) |
| Perimenopausal | 0 (0%) | 4 (1·6%) | 0 (0%) | 0 (0%) | 20 (6·8%) | 0 (0%) |
| Postmenopausal | 0 (0%) | 128 (51·8%) | 0 (0%) | 245 (47·4%) | 110 (37·7%) | 0 (0%) |
| Missing | 356 (100%)** | 57 (23·1%) | 552 (100%) | 6 (1.2%) | 2 (0·7%) | 0 (0%) |
| **Disease Stage** | | | | | | |
| I | 44 (12·4%) | 41 (16·6%) | 0 (0%) | 55 (10·6%) | 0 (0%) | 232 (59·5%) |
| II | 204 (57·3%) | 85 (34·4%) | 0 (0%) | 352 (68·1%) | 201 (68·8%) | 157 (40·3%) |
| III | 71 (19·9%) | 41 (16·5%) | 552 (100%) | 109 (21·1%) | 91 (31·2%) | 0 (0%) |
| IV | 10 (2·8%) | 2 (0·8%) | 0 (0%) | 0 (0%) | 0 (0%) | 0 (0%) |
| Missing | 27 (7·6%)** | 78 (31·6%) | 0 (0%) | 1 (0·2%) | 0 (0%) | 1 (0·3%) |
| **T-stage** | | | | | | |
| T0* | 0 (0%) | 0 (0%) | 0 (0%) | 0 (0%) | 1 (0·3%) | 0 (0%) |
| T1 | 0 (0%) | 59 (23·9%) | 129 (23·4%) | 236 (45·6%) | 17 (5·8%) | 232 (59·5%) |
| T2 | 0 (0%) | 92 (37·2%) | 328 (59·4%) | 249 (48·2%) | 186 (63·7%) | 151 (38·7%) |
| T3 | 0 (0%) | 20 (8·1%) | 85 (15·4%) | 29 (5·6%) | 75 (25·7%) | 6 (1·5%) |
| T4 | 0 (0%) | 8 (3·2%) | 0 (0%) | 2 (0·4%) | 12 (4·1%) | 0 (0%) |
| Tx | 0 (0%) | 68 (27·5%) | 0 (0%) | 0 (0%) | 1 (0·3%) | 0 (0%) |
| Missing | 356 (100%**) | 0 (0%) | 10 (1·8%) | 1 (0·2%) | 0 (0%) | 1 (0·3%) |
| **N-stage** | | | | | | |
| N0 | 0 (0%) | 83 (33·6%) | 0 (0%) | 107 (20·7%) | 114 (39·0%) | 390 (100%) |
| N1 | 0 (0%) | 58 (23·5%) | 0 (0%) | 315 (60·9%) | 138 (47·3%) | 0 (0%) |
| N2 | 0 (0%) | 25 (10·1%) | 0 (0%) | 74 (14·3%) | 21 (7·2%) | 0 (0%) |
| N3 | 0 (0%) | 4 (1·6%) | 0 (0%) | 21 (4·1%) | 19 (6·5%) | 0 (0%) |
| NX | 0 (0%) | 77 (31·2%) | 0 (0%) | 0 (0%) | 0 (0%) | 0 (0%) |
| Missing | 356 (100%)** | 0 (0%) | 552 (100%) | 0 (0%) | 0 (0%) | 0 (0%) |
| **M-stage** | | | | | | |
| M0 | 0 (0%) | 73 (29·6%) | 552 (100%) | 517 (100%) | 290 (99·3%) | 390 (100%) |
| M1 | 0 (0%) | 2 (0·8%) | 0 (0%) | 0 (0%) | 0 (0%) | 0 (0%) |
| MX | 0 (0%) | 172 (69·6%) | 0 (0%) | 0 (0%) | 2 (0·7%) | 0 (0%) |
| Missing | 356 (100%)** | 0 (0%) | 0 (0%) | 0 (0%) | 0 (0%) | 0 (0%) |
| **Tumour grade** | | | | | | |
| 1 | 0 (0%) | 32 (13·0%) | 121 (21·9%) | 53 (10·3%) | 12 (4·1%) | 2 (0·5%) |
| 2 | 0 (0%) | 90 (36·4%) | 207 (37·5%) | 220 (42·6%) | 138 (47·3%) | 52 (13·3%) |
| 3 | 0 (0%) | 124 (50·2%) | 205 (37·1%) | 221 (42·7%) | 125 (42·8%) | 336 (86·2%) |
| Missing | 356 (100%)** | 1 (0·4%) | 19 (3·4%) | 23 (4·4%) | 17 (5·8%) | 0 (0%) |
| **Lymphovascular invasion** | | | | | | |
| Absent | 0 (0%) | 182 (73·7%) | 0 (0%) | 0 (0%) | 0 (0%) | 349 (89·5%) |
| Present | 0 (0%) | 40 (16·2%) | 0 (0%) | 0 (0%) | 0 (0%) | 40 (10·3%) |
| Missing | 356 (100%)** | 25 (10·1%) | 552 (100%) | 517 (100%) | 292 (100%) | 1 (0·3%) |
| **Breast cancer subtype** | | | | | | |
| ER+/HER2+ | 32 (9·0%) | 3 (1·2%) | 0 (0%) | 8 (1·5%) | 187 (64·0%) | 0 (0%) |
| ER+/HER2- | 168 (47·2%) | 188 (76·1%) | 410 (74·3%) | 409 (79·1%) | 0 (0%) | 0 (0%) |
| ER-/HER2+ | 9 (2·5%) | 2 (0·8%) | 0 (0%) | 7 (1·4%) | 105 (36·0%) | 0 (0%) |
| TN | 112 (31·5%) | 54 (21·9%) | 142 (25·7%) | 92 (17·8%) | 0 (0%) | 390 (100%) |
| Missing | 35 (9·8%)** | 0 (0%) | 0 (0%) | 1 (0·2%) | 0 (0%) | 0 (0%) |
| **Histology** | | | | | | |
| BC NST | 321 (90·2%) | 193 (78·1%) | 115 (20·8%) | 414 (80·1%) | 265 (90·8%) | 333 (85·4%) |
| ILC | 18 (5·1%) | 17 (6·9%) | 50 (9·1%) | 71 (13·7%) | 13 (4·5%) | 1 (0·3%) |
| Metaplastic | 0 (0%) | 3 (1·2%) | 0 (0%) | 0 (0%) | 0 (0%) | 20 (5·1%) |
| Other | 3 (0·8%) | 34 (13·8%) | 39 (7·1%) | 25 (4·8%) | 14 (4·8%) | 36 (9·2%) |
| Missing | 14 (3·9%)** | 0 (0%) | 348 (63·0%) | 7 (1·4%) | 0 (0%) | 0 (0%) |
| **TILs score pathologist (%)** | | | | | | |
| Median [Min, Max] | 12·5 [1·00, 90·0] | 13·0 [0, 93·0] | 15·0 [0, 90·0] | 10·0 [0·500, 77·5] | 5·00 [1·00, 92·5] | 20·0 [1·00, 95·0] |
| Missing | 14 (3.9%)** | 0 (0%) | 0 (0%) | 0 (0%) | 0 (0%) | 0 (0%) |

Table 1: Baseline characteristics of all clinical cohorts. ER: Estrogen receptor; HER2: Human Epidermal growth factor Receptor 2, TN: Triple Negative; BC NST: Breast Cancer of No Special Type; ILC: Invasive Lobular Carcinoma; TILs: Tumour Infiltrating Lymphocytes.

*Patient with no evidence of tumour in the breast but cancer has spread to axillary lymph nodes.

**For TCGA, 14 patients have two available slides, which are included in the sample count (n=356). In the table, we present patient characteristics; hence, the clinical data is presented as missing for 14 samples.

**Table 2: Cox regression results**

| | No TILs | | Pathologist TILs | | ECTIL-TCGA | | ECTIL-combined | | ECTIL-TNBC | |
|---|---|---|---|---|---|---|---|---|---|---|
| | N= 384 | E=139 | N = 384 | E = 139 | N= 384 | E=139 | N= 384 | E= 139 | N= 384 | E= 139 |
| **TILs Univariable** | | | | | | | | | | |
| Concordance Index | | | 0·62 (p<0·001) | | 0·57 (p<0·01) | | 0·58 (p<0·001) | | 0·58 (p<0·001) | |
| Variable | HR | 95% CI | HR | 95% CI | HR | 95% CI | HR | 95% CI | HR | 95% CI |
| TILs | | | | | | | | | | |
| 10% increment | | | 0·87 | 0·82 to 0·92, **P<0·001** | 0·87 | 0·79 to 0·96, **p<0·01** | 0·86 | 0·78 to 0·94, **p<0·01** | 0·85 | 0·77 to 0·94, **p<0·01** |
| **TILs Multivariable** | | | | | | | | | | |
| Concordance index | 0·63 (p<0·001) | | 0·68 (p<0·001 | | 0·65 (p<0·001) | | 0·65 (p<0·001) | | 0·65 (p<0·001) | |
| Variable | HR | 95% CI | HR | 95% CI | HR | 95% CI | HR | 95% CI | HR | 95% CI |
| TILs | | | | | | | | | | |
| 10% increment | | | 0·86 | 0·81 to 0·92, **p<0·001** | 0·87 | 0·79 to 0·95, **p<0·01** | 0·85 | 0·77 to 0·93, **p<0·001** | 0·84 | 0·76 to 0·93, **p<0·001** |
| T-stage | | | | | | | | | | |
| T1a/b (ref) | 1 | | 1 | | 1 | | 1 | | 1 | |
| T1c | 1·02 | 0·55 to 1·91 | 1·02 | 0·55 to 1·91 | 1·11 | 0·59 to 2·09 | 1·09 | 0·58 to 2·05 | 1·11 | 0·59 to 2·09 |

| | | | | | | | | | | | |
|---|---|---|---|---|---|---|---|---|---|---|---|
| T2/3 | 1·50 | 0·81 to 2·80 | 1·62 | 0·87 to 3·02 | 1·62 | 0·87 to 3·02 | 1·62 | 0·87 to 3·02 | 1·66 | 0·89 to 3·10 | |
| Grade | | | | | | | | | | | |
| Grade 1 or 2 (ref) | 1 | | 1 | | 1 | | 1 | | 1 | | |
| Grade 3 | 0·82 | 0·52 to 1·28 | 1·01 | 0·64 to 1·60 | 0·88 | 0·56 to 1·39 | 0·92 | 0·58 to 1·46 | 0·90 | 0·57 to 1·43 | |
| Histology | | | | | | | | | | | |
| BC NST (ref) | 1 | | 1 | | 1 | | 1 | | 1 | | |
| Metaplastic | 0·34 | 0·11 to 1·08 | 0·25 | 0·08 to 0·81, **p<0·05** | 0·32 | 0·10 to 1·03 | 0·32 | 0·10 to 1·03 | 0·32 | 0·10 to 1·02 | |
| Others | 0·61 | 0·32 to 1·16 | 0·62 | 0·32 to 1·20 | 0·60 | 0·31 to 1·15 | 0·62 | 0·32 to 1·18 | 0·62 | 0·32 to 1·18 | |
| Lymphovascular invasion | | | | | | | | | | | |
| Absent (ref) | 1 | | 1 | | 1 | | 1 | | 1 | | |
| Present | 2·05 | 1·30 to 3·23, **p<0·01** | 1·87 | 1·18 to 2·97, **p<0·01** | 1·98 | 1·25 to 3·14, **p<0·01** | 2·05 | 1·29 to 3·24, **p<0·01** | 2·00 | 1·26 to 3·17, **p<0·01** | |
| Local treatment | | | | | | | | | | | |
| BCS + radiotherapy (ref) | 1 | | 1 | | 1 | | 1 | | 1 | | |
| Mastectomy | 1·39 | 0·96 to 2·00 | 1·41 | 0·98 to 2·04 | 1·47 | 1·02 to 2·12, **p<0·05** | 1·47 | 1·02 to 2·13, **p<0·05** | 1·48 | 1·03 to 2·14, **p<0·05** | |
| Other | 1·91 | 0·96 to 2·00 | 1·76 | 0·71 to 4·36 | 1·81 | 0·73 to 4·49 | 1·93 | 0·78 to 4·79 | 1·85 | 0·75 to 4·60 | |

Table 2: Cox regression results. Uni and multivariable Cox regression results, comparing the pathologist's score and the various ECTIL models that are trained on different (sub)sets of data (ECTIL-TCGA, ECTIL-combined, ECTIL-TNBC). For both the univariable and multivariable regression models we find that all computational models achieve a similar hazard ratio to the pathologist's score. N: number of patients; E: number of events; HR: Hazard ratio; 95% CI: 95% confidence intervals; ref: reference; TILs: Tumour Infiltrating Lymphocytes; BC NST: Breast Cancer of No Special Type; BCS: Breast Conserving Surgery

## Figures

### Figure 1: model overview

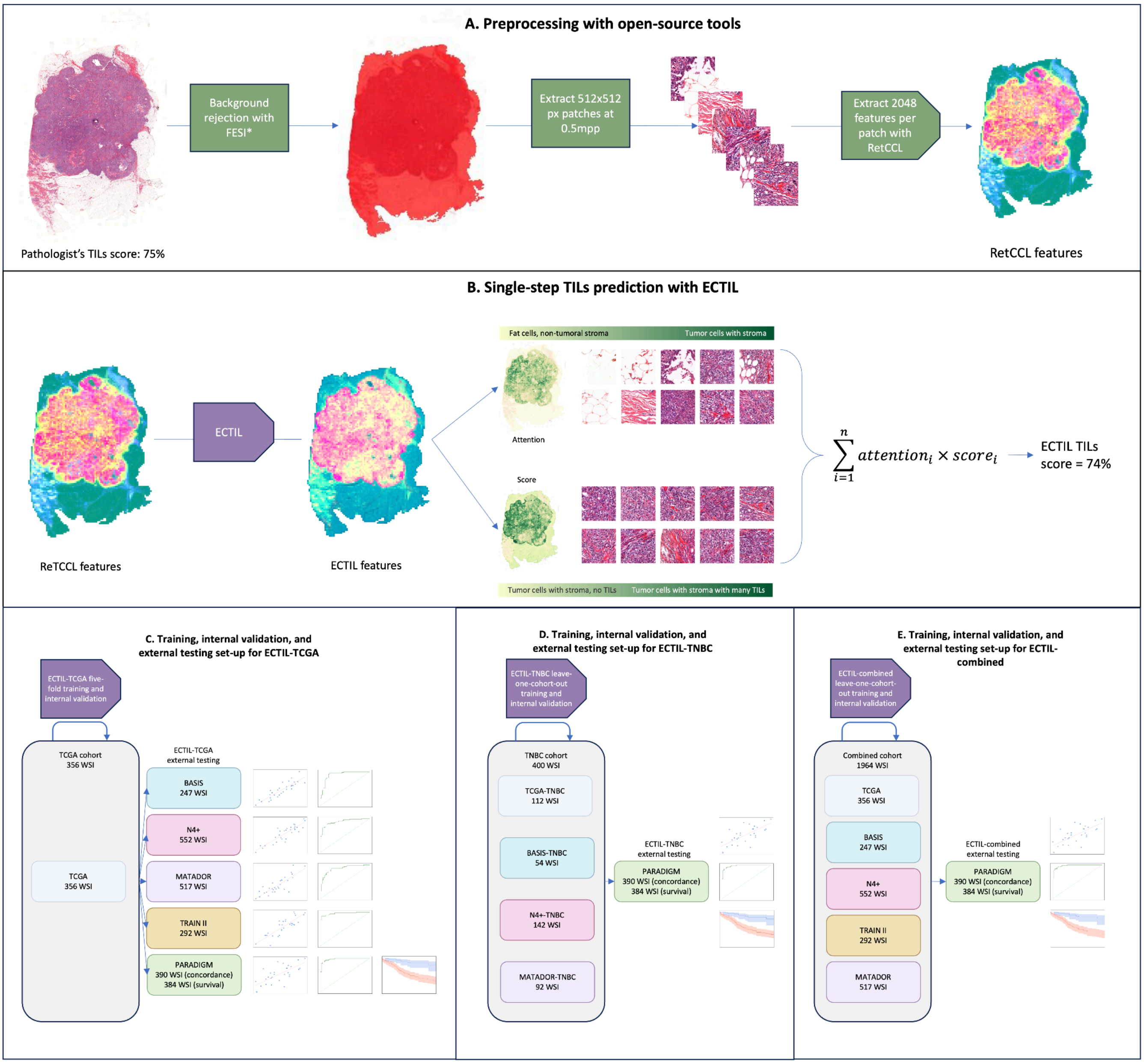

$$\sum_{i=1}^{n} attention_i \times score_i$$

Overview of data preprocessing (see tissue selection and feature extraction sections), training, internal validation, and external testing set-up (see model training and evaluation section), and the single-step inference of ECTIL (see model section).

A) For data preprocessing, we use FESI to reject the background, after which we extract 512x512 px tiles at 0·5 microns per pixel. Using RetCCL, we extract 2048 features for each tile.

B) The single-step ECTIL inference takes as input the RetCCL features of a WSI that are mapped to a 512-dimensional feature space, which is used to compute the attention and the score for each tile. For both the attention and the score, we show the heatmap for this sample, and pick random tiles over the score and attention distribution and describe the content of these patches. Finally, the attention-weighted mean score of all tiles is the final model prediction.

For training, validation, and testing, we train three models. The scatter plots, ROC curves, and survival curves are generated with synthetic data for illustration purposes only.

C) ECTIL-TCGA (five-fold training and validation on TCGA), tested on all other cohorts.
D) ECTIL-TNBC (leave-one-cohort-out training and validation on the TNBC samples of all collected cohorts except PARADIGM), tested on PARADIGM.
E) ECTIL-combined (leave-one-cohort-out training and validation on all samples of all collected cohorts except PARADIGM).

For all external test sets, we analyse concordance with the pathologist using the Pearson's r and AUROC for TILs-high versus TILs-low prediction. For the PARADIGM cohort we additionally perform overall survival analyses.

**Figure 2: Overview of performance of all models on all datasets**

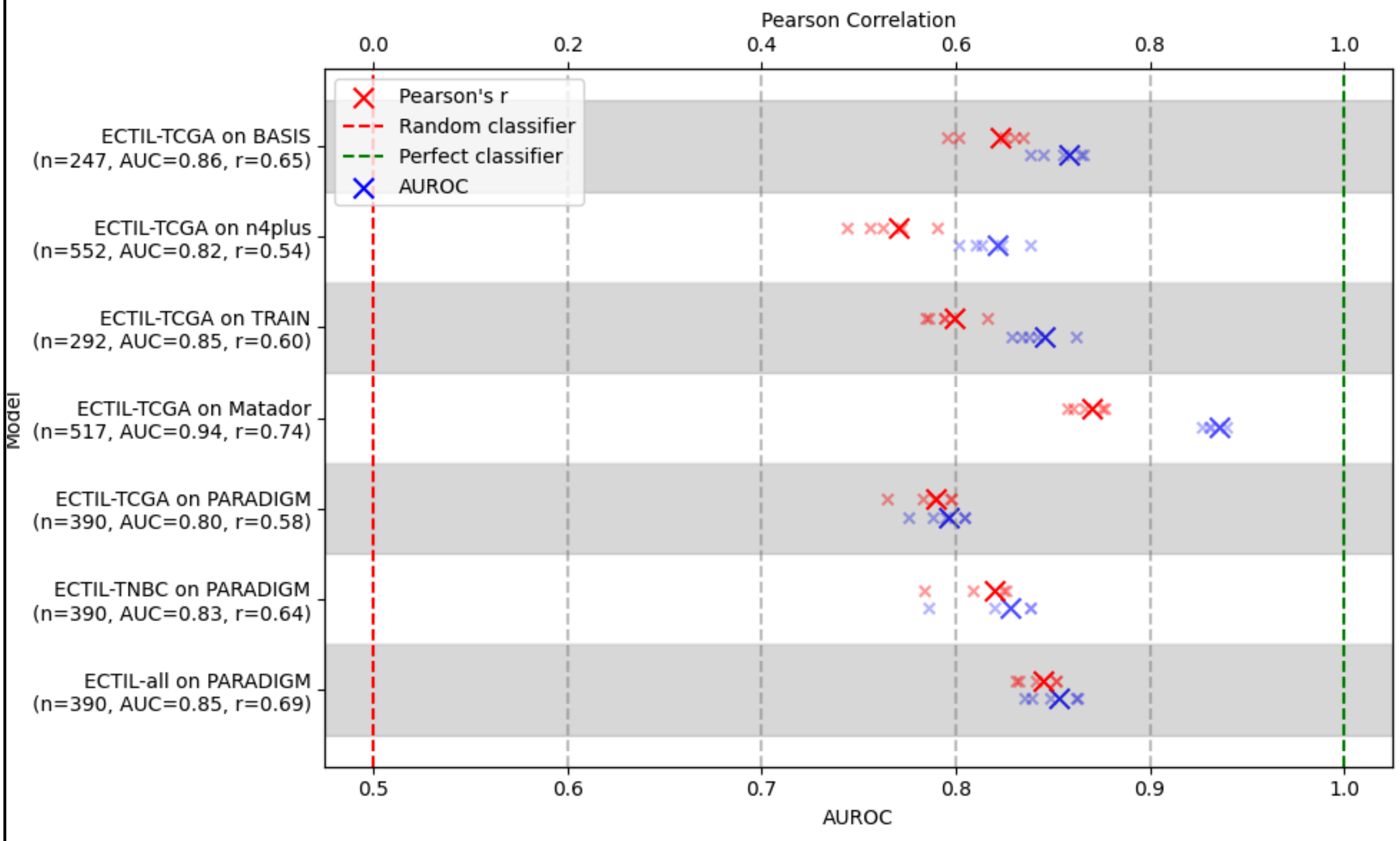

Overview of external test metrics from all performed experiments. The large marks represent the metric for the ensemble which is used for further analysis. The transparent small marks are the results of each separate model that has been trained on various folds, to show how sensitive the predictions are to various training data. The bottom x-axis displays the area under the receiver operating characteristic curve, where the pathologist score is cut at an TILs score of 30 to generate TILs-high and TILs-low labels. The top x-axis displays the Pearson's correlation coefficient. The y-axis displays the test results of various models on various cohorts. A random model would achieve a Pearson's correlation coefficient of 0·0 and an AUROC of 0·5, whereas a perfect model would achieve a pearson's correlation coefficient of 1·0 and AUROC of 1·0.

**Figure 3: Calibration curve of ECTIL-combined on PARADIGM**

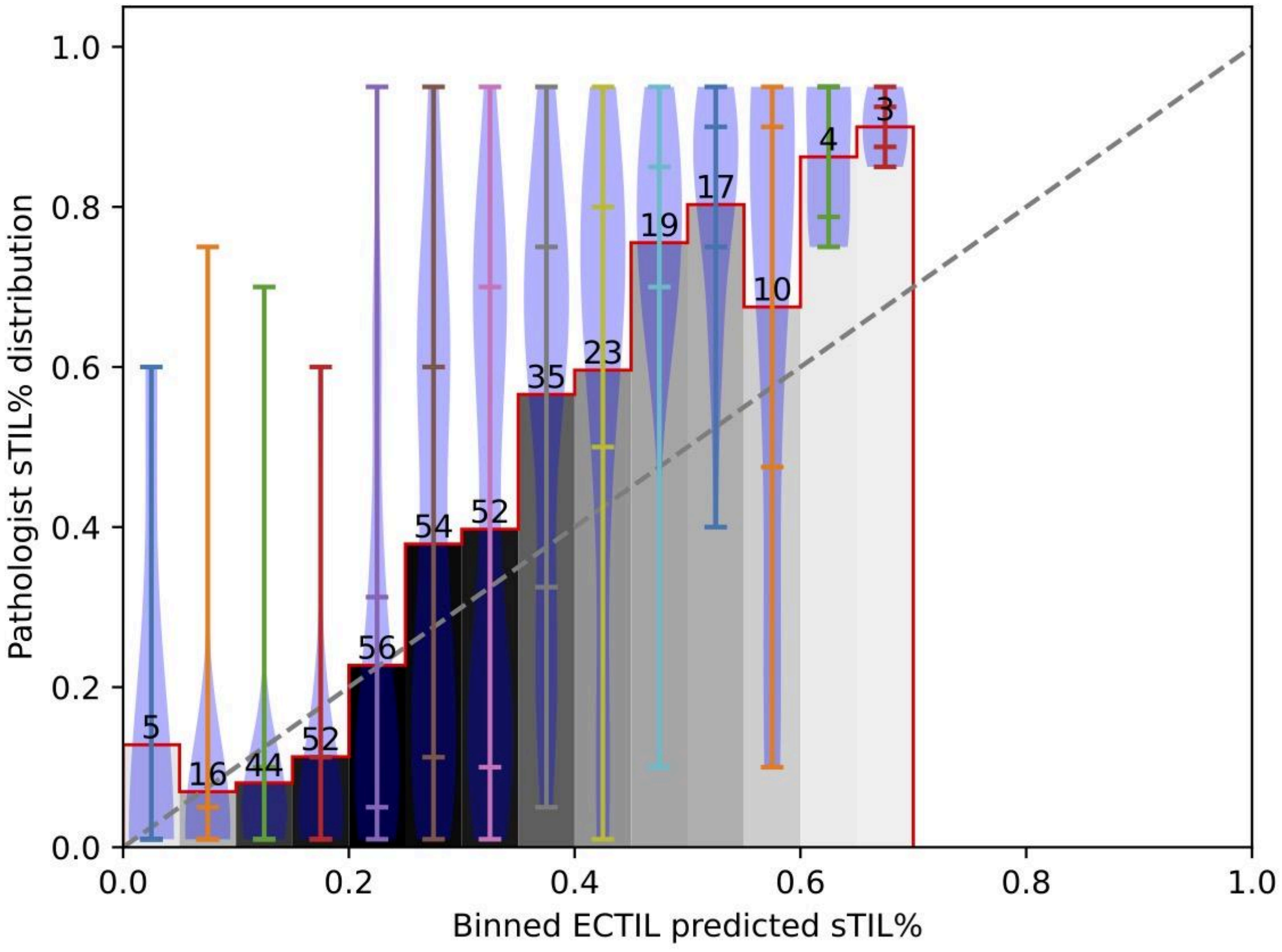

Figure 3: This figure displays a calibration curve of the ECTIL model trained on all cohorts which is validated on the PARADIGM cohort. The curve visualises the pathologist score distribution conditional on model predicted scores $p(y|\hat{y})$, where y is the ground-truth score, and $\hat{y}$ is a bin of estimated scores. A perfectly predictive and perfectly calibrated model would have the bars located exactly on the $x = y$ line over the entire distribution. On the x-axis we see 20 uniform bins of ECTIL predictions, while the y-axis shows the pathologist score, as scored by a (panel of) pathologist(s) for those samples that were predicted to be inside that bin by ECTIL. The colour of each bar represents the number of samples in the bin (the bin with the most samples is black and the bin with the least samples is white), and the exact numbers are displayed above the bar. The violin plots for each bin show the distribution of pathologist scores with the min, 10th percentile, 90th percentile, and max, while the bar plot displays the mean score for that bin.

**Supplementary materials**

**Supplementary sections**

**Section S1: Ethical considerations**

All clinical trials in this study are approved by the ethical committee of the Netherlands Cancer Institute. All studies are conducted according to the Good Clinical Practice guidelines and the Declaration of Helsinki. For all randomised trials, patients provide written consent for study treatment and use of tumour tissue for the purpose of translational research.

The current study and data collection was approved by the Institutional Review Board of the Netherlands Cancer Institute.

**Section S2: Clinical trial description**

N4+ (NCT03087409) is a multicentre, randomised, phase III trial for patients with stage III breast cancer (no distant metastasis but at least four tumour positive axillary lymph nodes). Between 1993 and 1999, 855 patients were randomised (1:1) to receive either adjuvant high dose chemotherapy or conventional dose chemotherapy. As the trial started before anti-HER2-targeted therapy, many of the follow-up studies exclude the HER2+ patients, which is why in our current study we only have WSI of the HER2- patients.[30] MATADOR (Microarray Analysis in breast cancer to Tailor Adjuvant Drugs Or Regimens) (ISRCTN registry - ISRCTN61893718) is a phase III, multicentre randomised trial for patients with stage II/III breast cancer. Between 2004 and 2012, 664 patients were randomised between 6 cycles of adjuvant dose dense doxorubicin and cyclophosphamide (ddAC) or adjuvant docetaxel, doxorubicin and cyclophosphamide (TAC).[28] The TRAIN II (NCT01996267), is a multicentre, randomised, phase III de-escalation trial for patients with stage II-III HER2+ breast cancer. Between 2013 and 2016, 438 patients were randomised (1:1) to neoadjuvant chemotherapy with or without the addition of anthracyclines in combination with trastuzumab.[29]

**Section S3: Hyperparameter search**

We perform a Bayesian hyperparameter search on a single randomly selected fold of the 5-fold data split of TCGA samples. All models are trained with the MSE loss, the ADAM optimizer, and L2 norm regularisation. The search was run in 80 trials using Hydra's Optuna Sweeper plugin on the learning rate (between 0·00001 and 0·01), weight decay (between 0·00001 and 0·01), post encoder size (128, 256, 512), batch size (8, 16, 32), attention hidden features (128, 256) and various dropout rates (parameter-wise and patch-wise for the post-encoder and classifier, probabilities of 0, 0·1, and 0·4). Smooth validation loss curves and high performance until 50 epochs (15 epochs patience on explained variance on the validation set) were used to hand-pick the optimal parameter set.

The final parameter set was chosen to be a learning rate of 0·0001, weight decay of 0·0006, post encoder projecting to 512 features with no dropout during training, a batch size of 16, attention layer hidden features of 128, and the final classifier with a feature-level dropout of 0·4 and a tile-level dropout of 0·1 during training.

**Supplementary figures**

**Figure S1: Inclusion Flow Diagram**

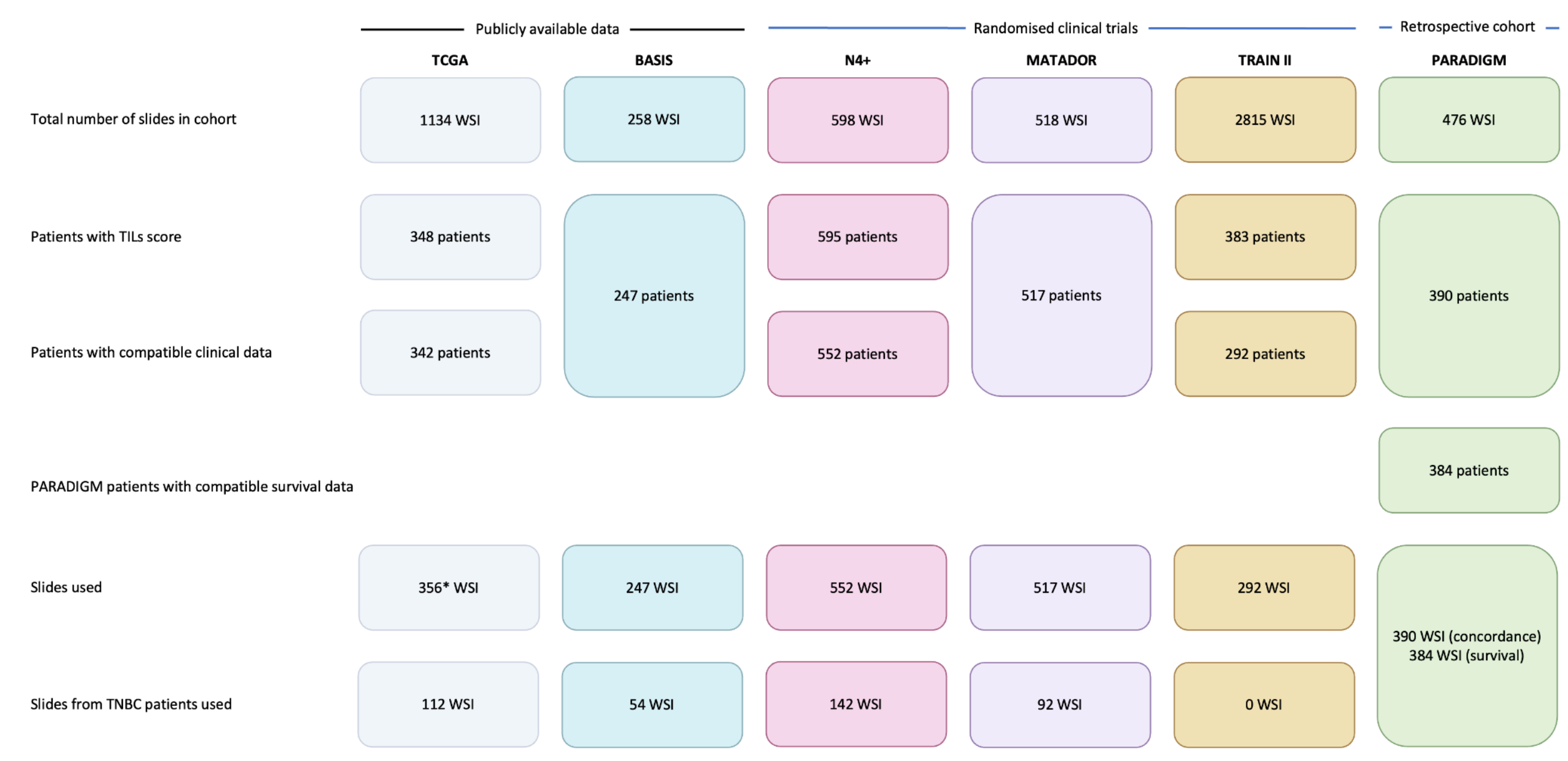

Figure S1: Inclusion flow diagram for all cohorts used in this study. TILs: Tumour Infiltrating Lymphocytes; WSI: Whole Slide Images; TNBC: Triple-negative breast cancer.

*For TCGA we include two slides from 14 patients as described in the methods section. Hence we have 356 slides for 342 patients.

**Figure S2: Scatter plots of ECTIL-TCGA tested on the internal held-out TCGA test set, and all external cohorts**

| | |
|---|---|
| TCGA | 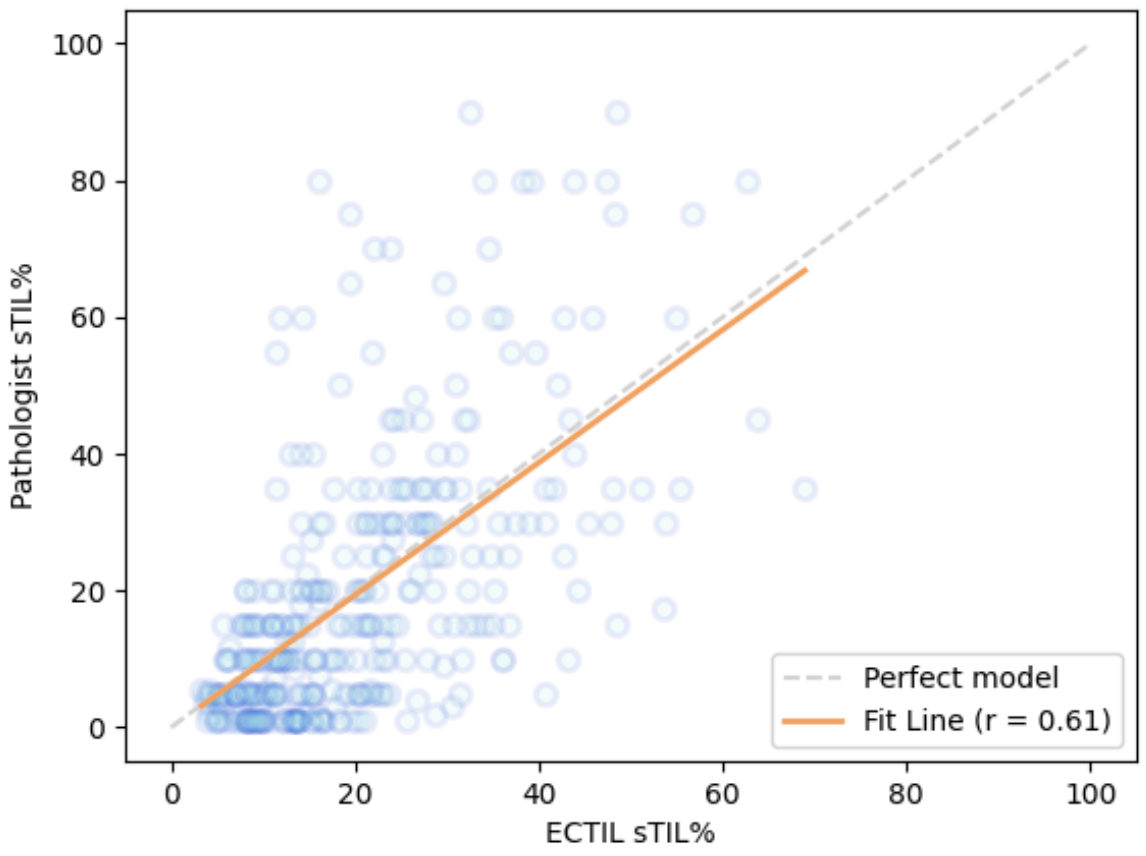 |
| BASIS | 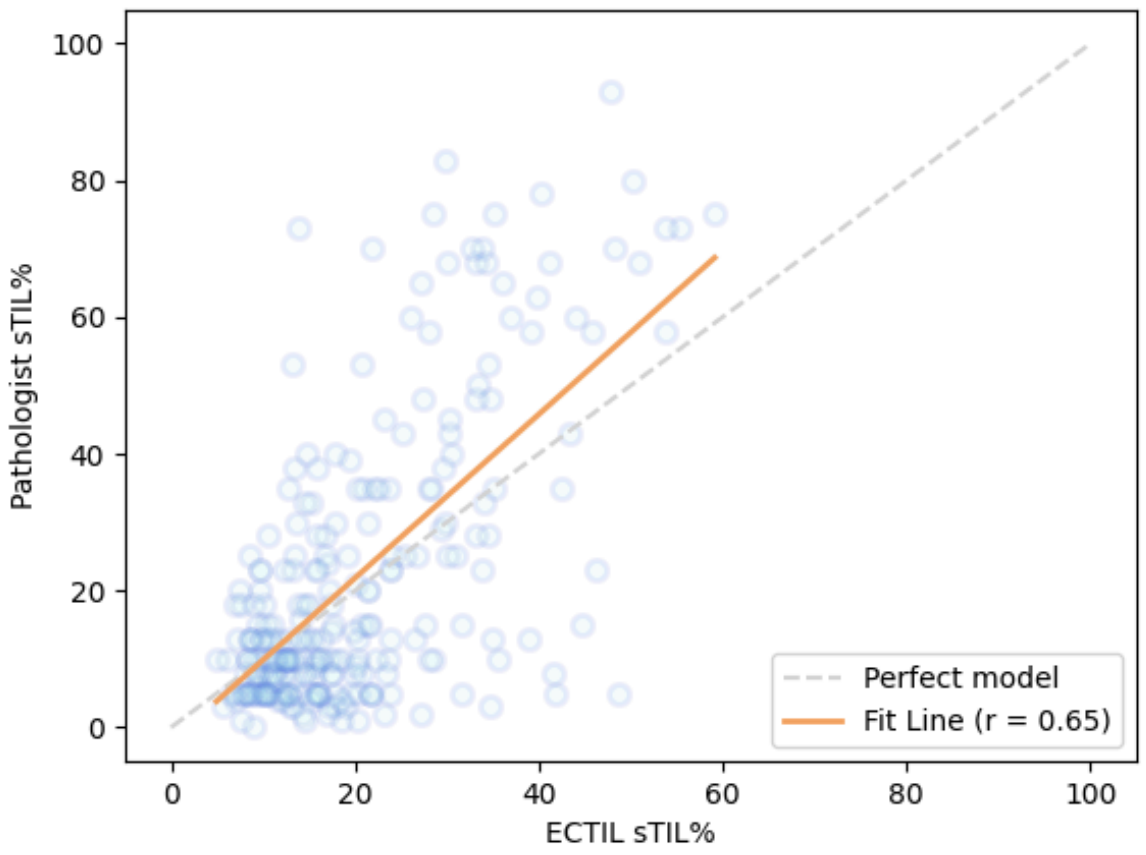 |
| N4+ | 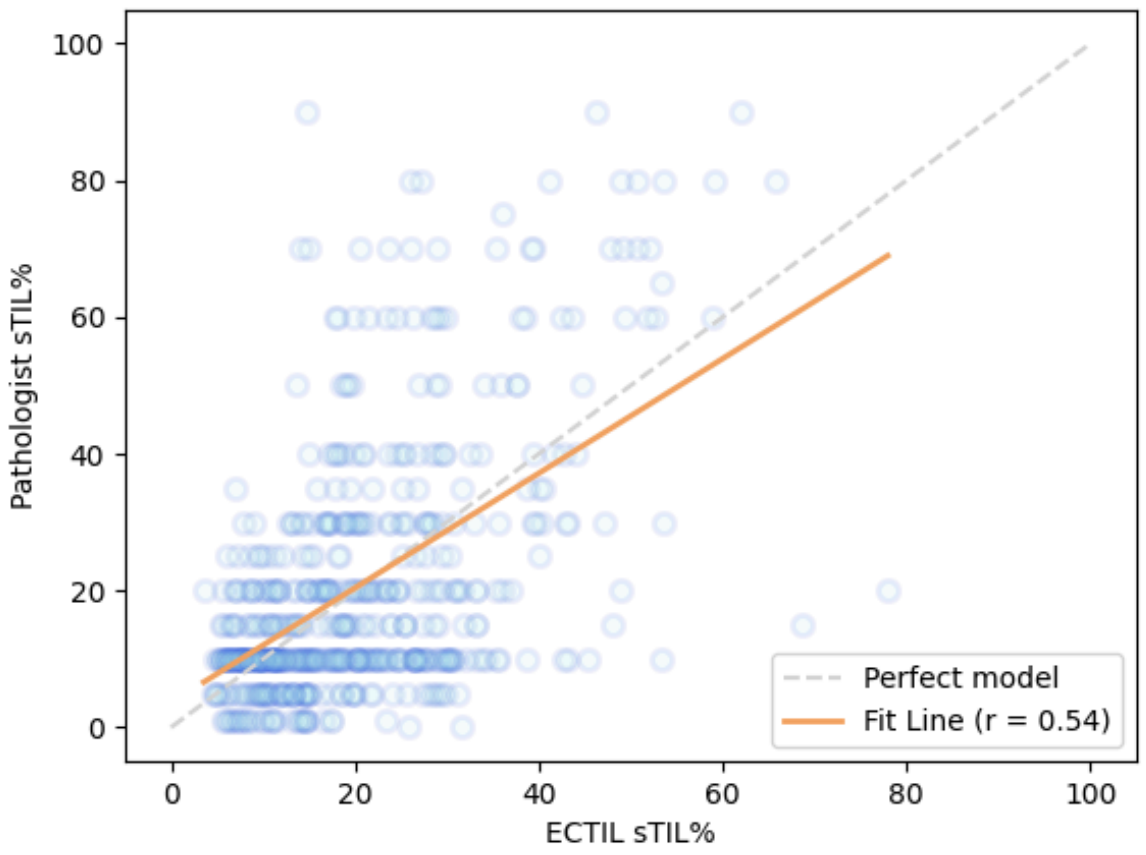 |

TRAIN

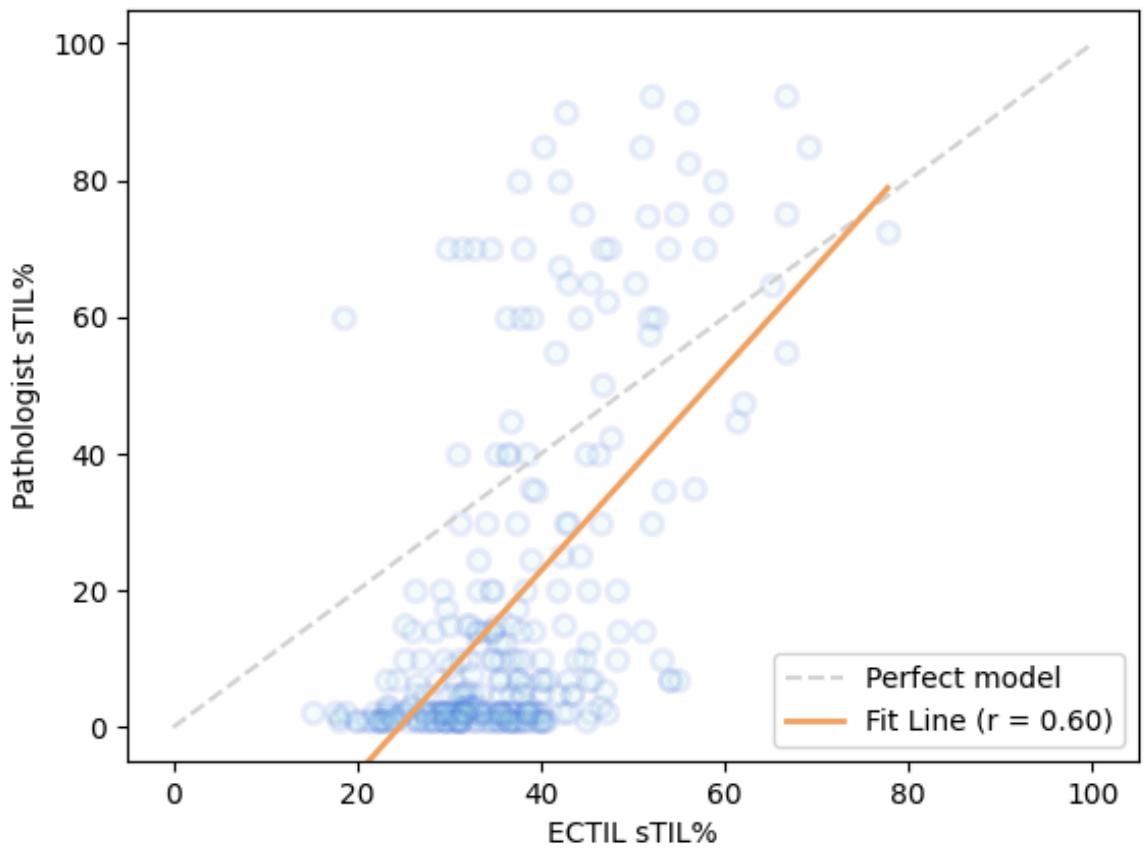

MATADOR

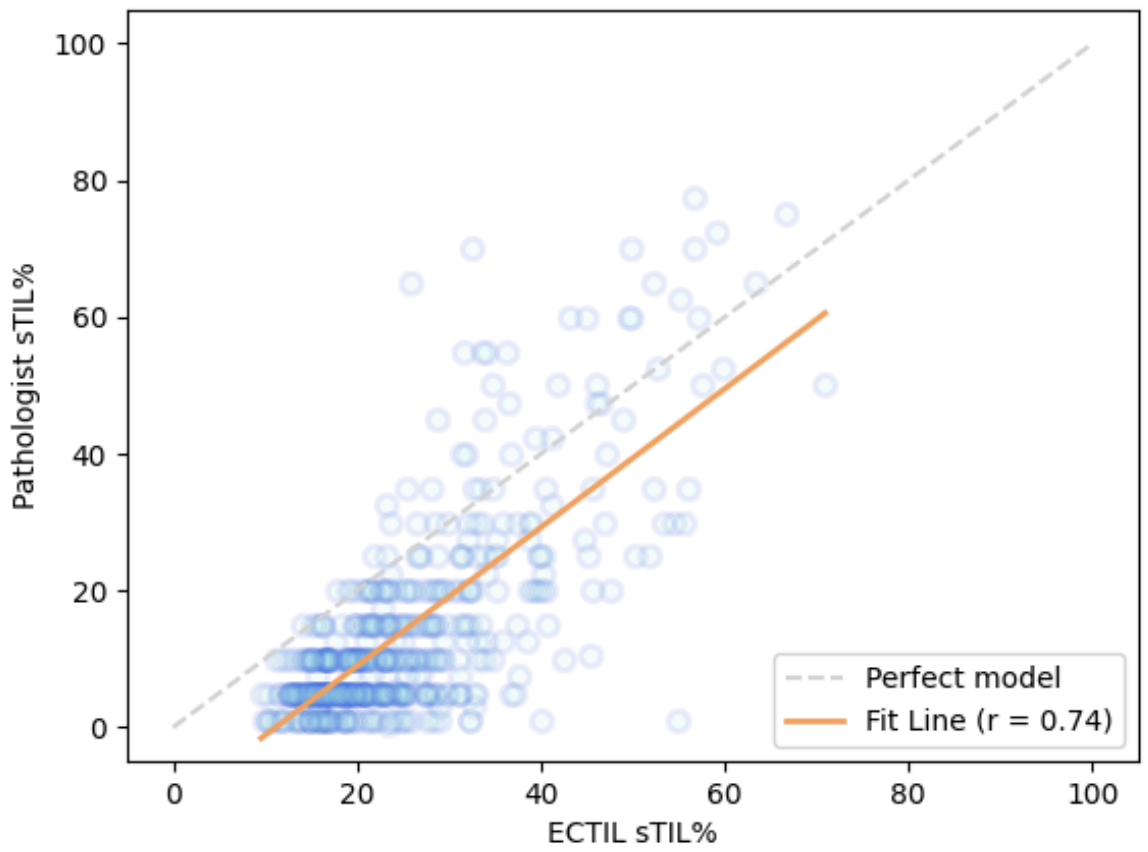

PARADIGM

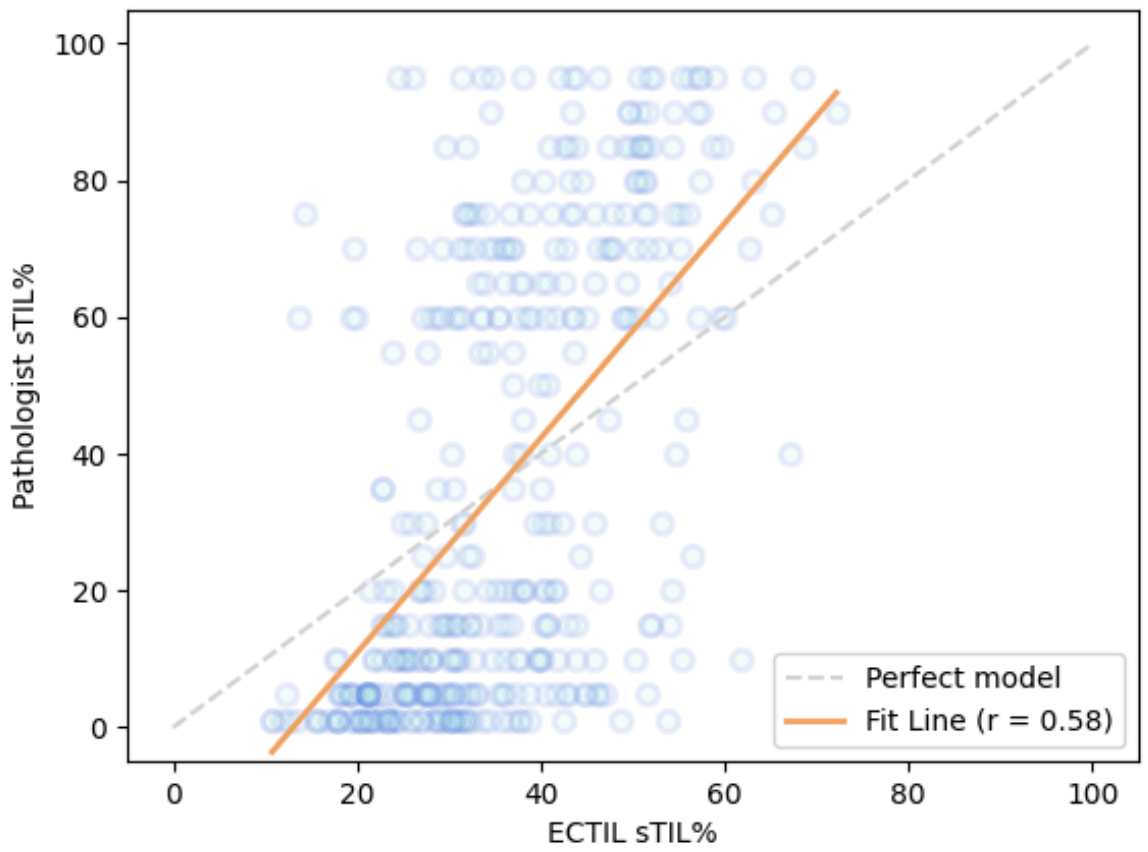

Results of ECTIL-TCGA on all external cohorts. The prediction for each patient is on a single slide, and the score is provided by averaging the prediction of the five models as trained on the five folds trained on TCGA.

**Figure S3: Calibration curve of ECTIL-TCGA on PARADIGM**

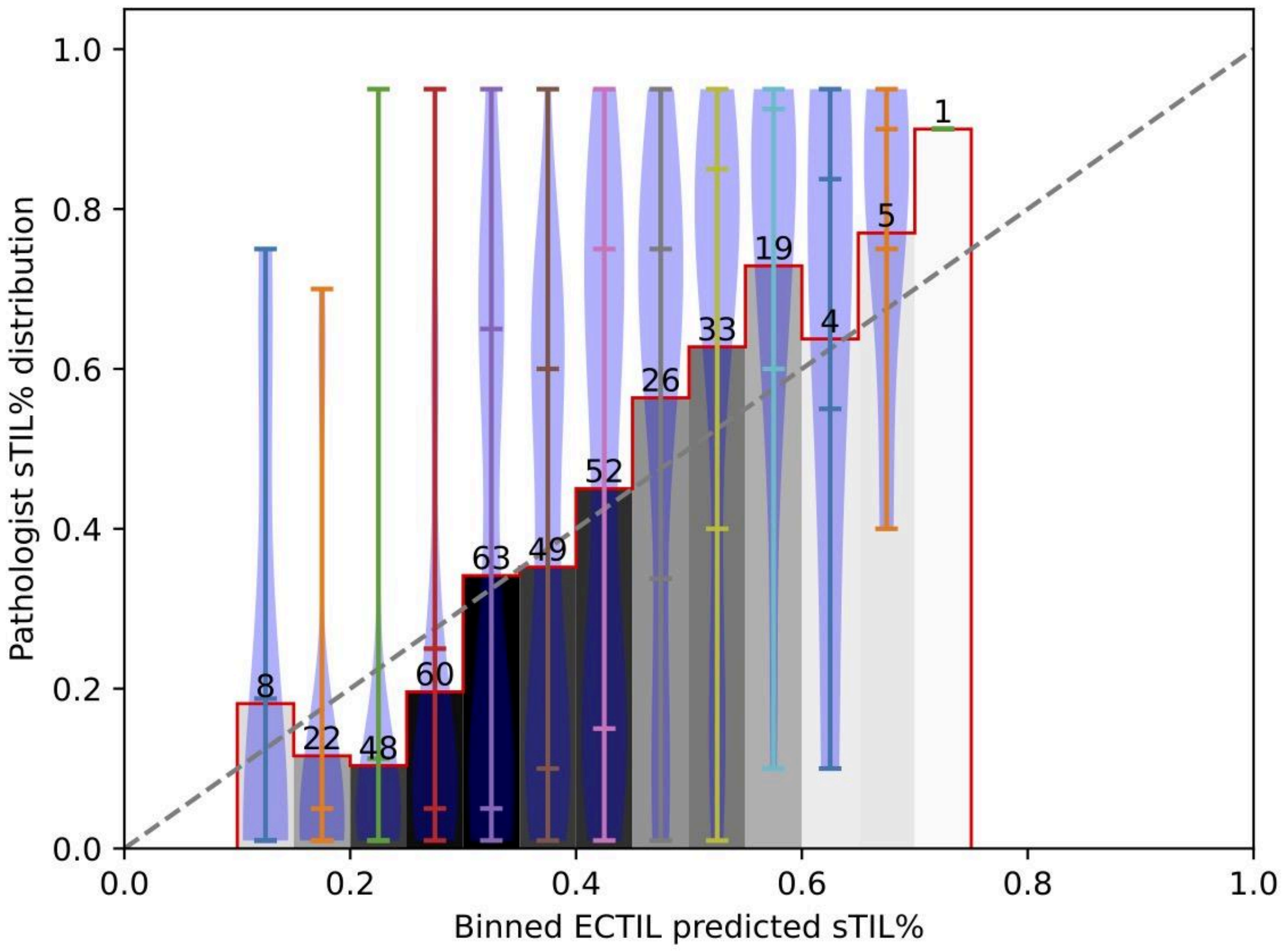

Calibration curve of ECTIL-TCGA on PARADIGM. See Figure 3 for the interpretation of a calibration curve.

**Figure S4: Calibration curve of ECTIL-TNBC on PARADIGM**

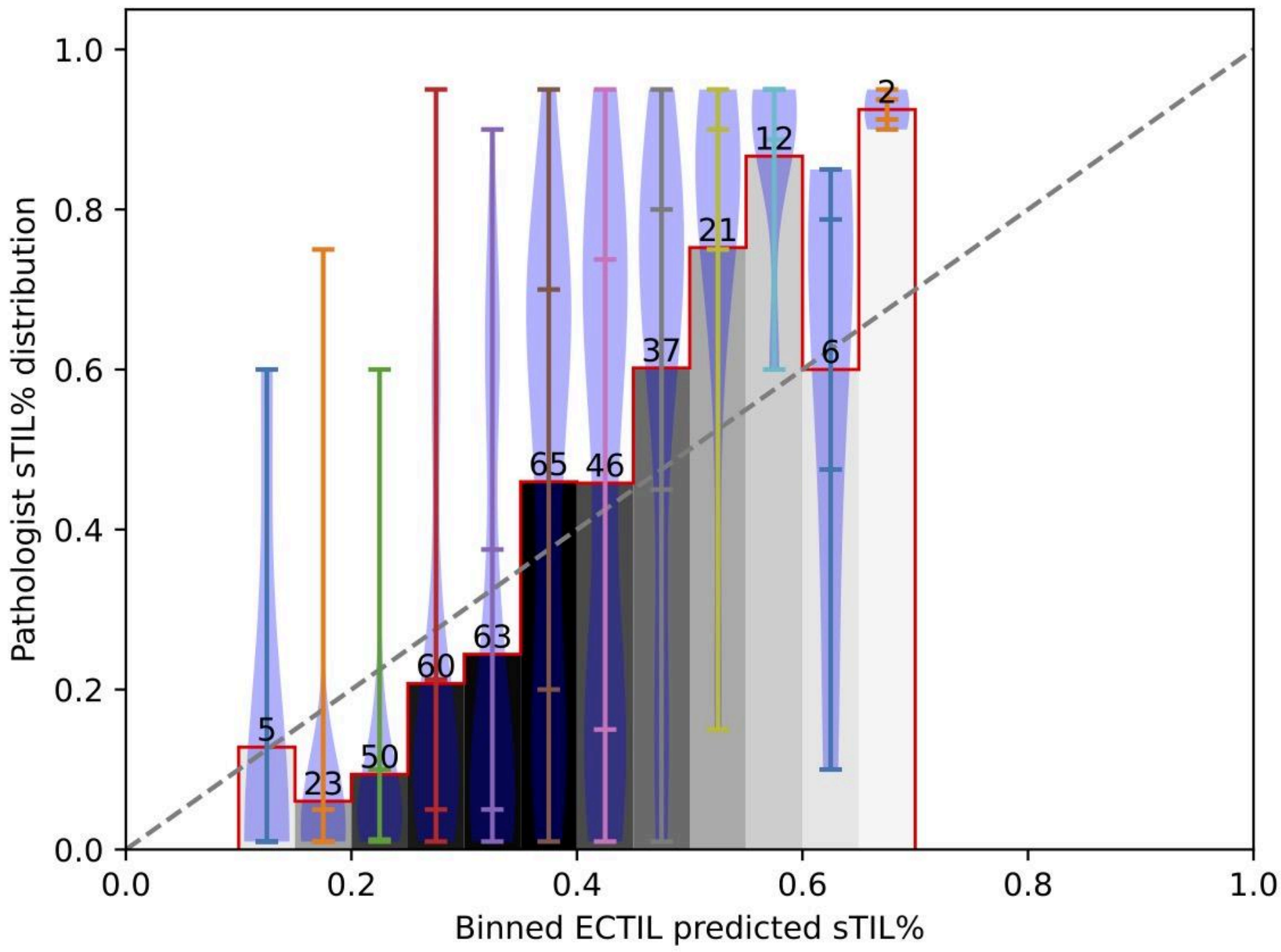

Calibration curve of ECTIL-TNBC tested on PARADIGM. See Figure 3 for the interpretation of a calibration curve.

**Figure S5-S7: Discordant cases**

**Figure S5-1: Discordant case of BASIS**

a. Thumbnail

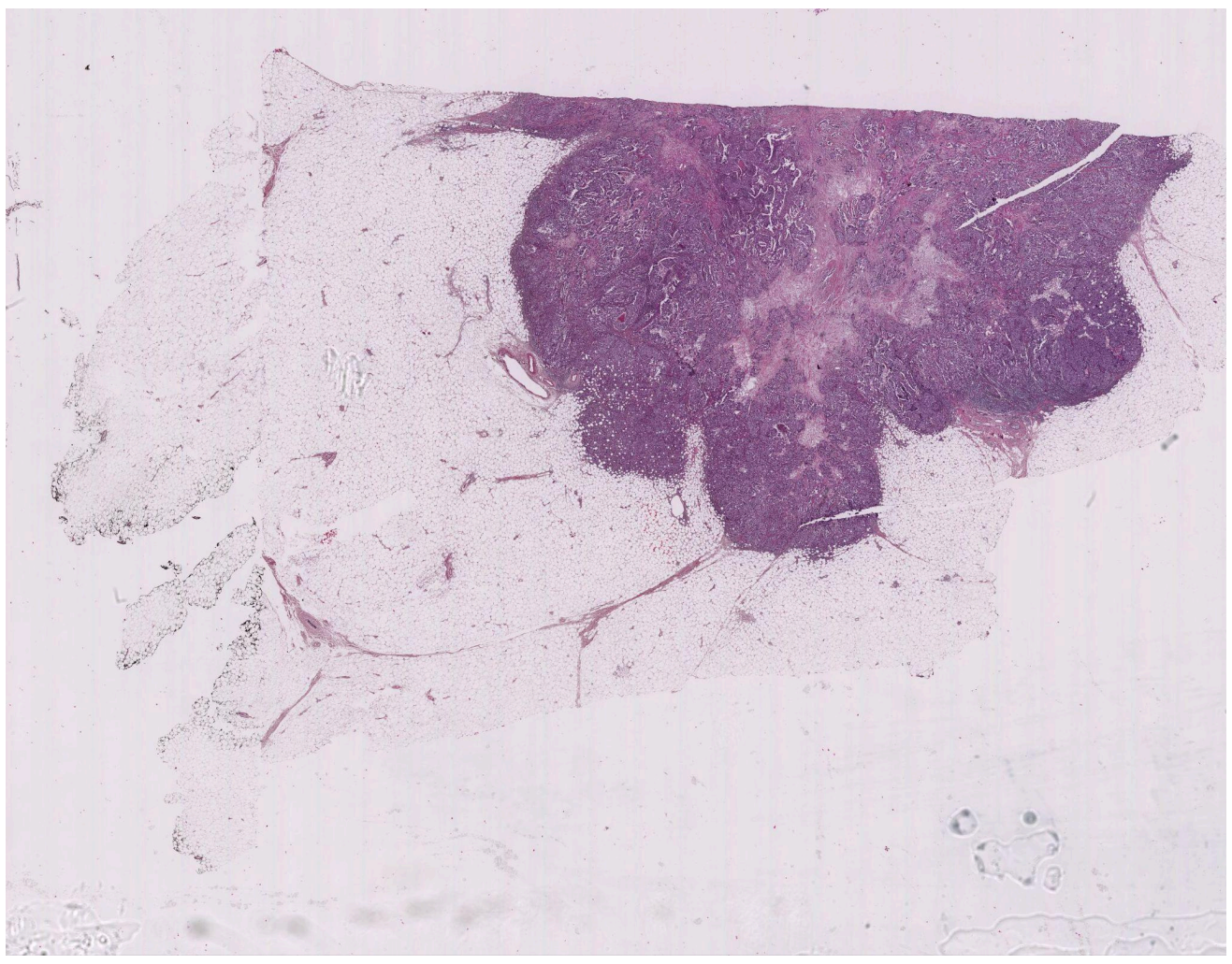

b. Tissue mask

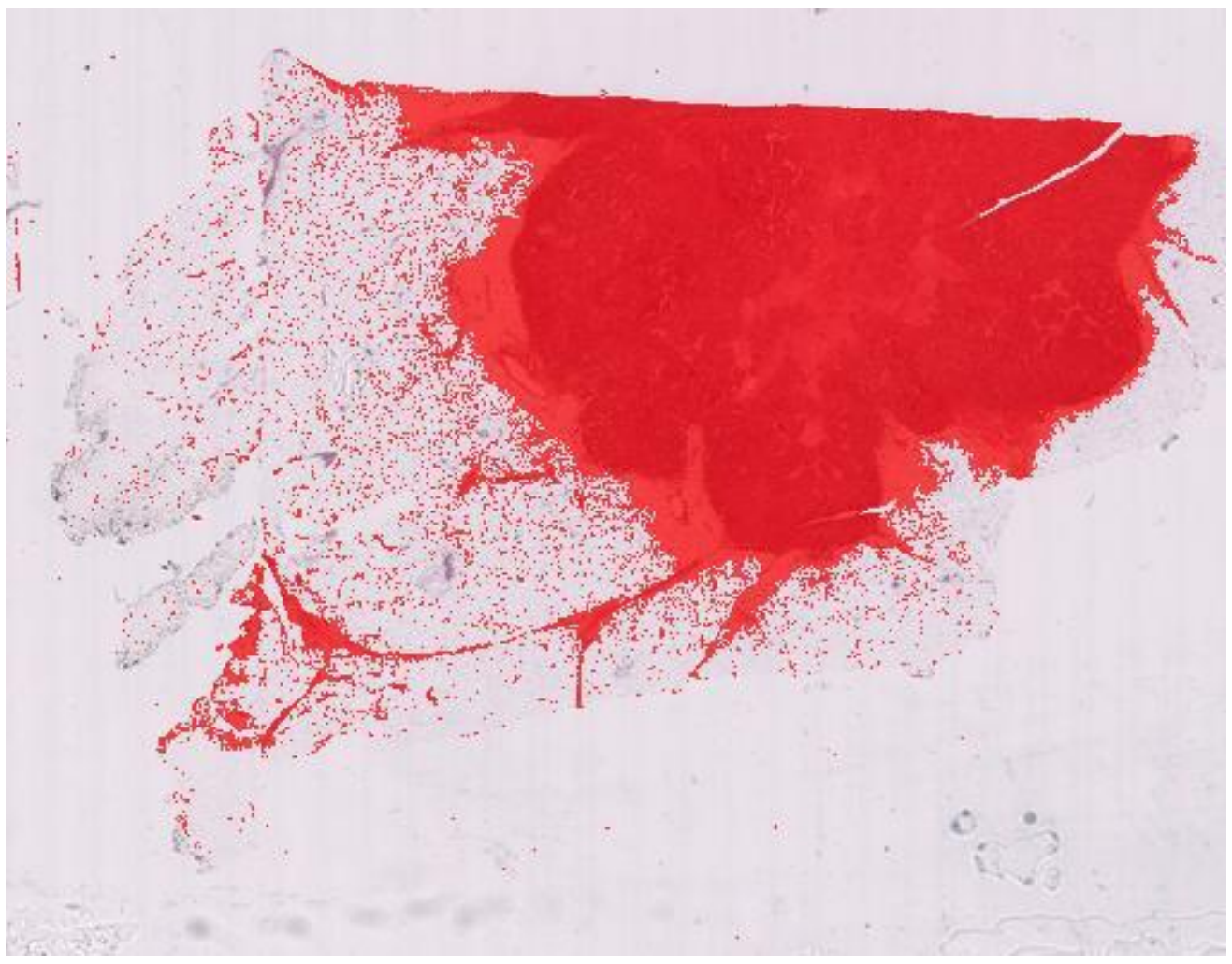

c. Attention heatmap

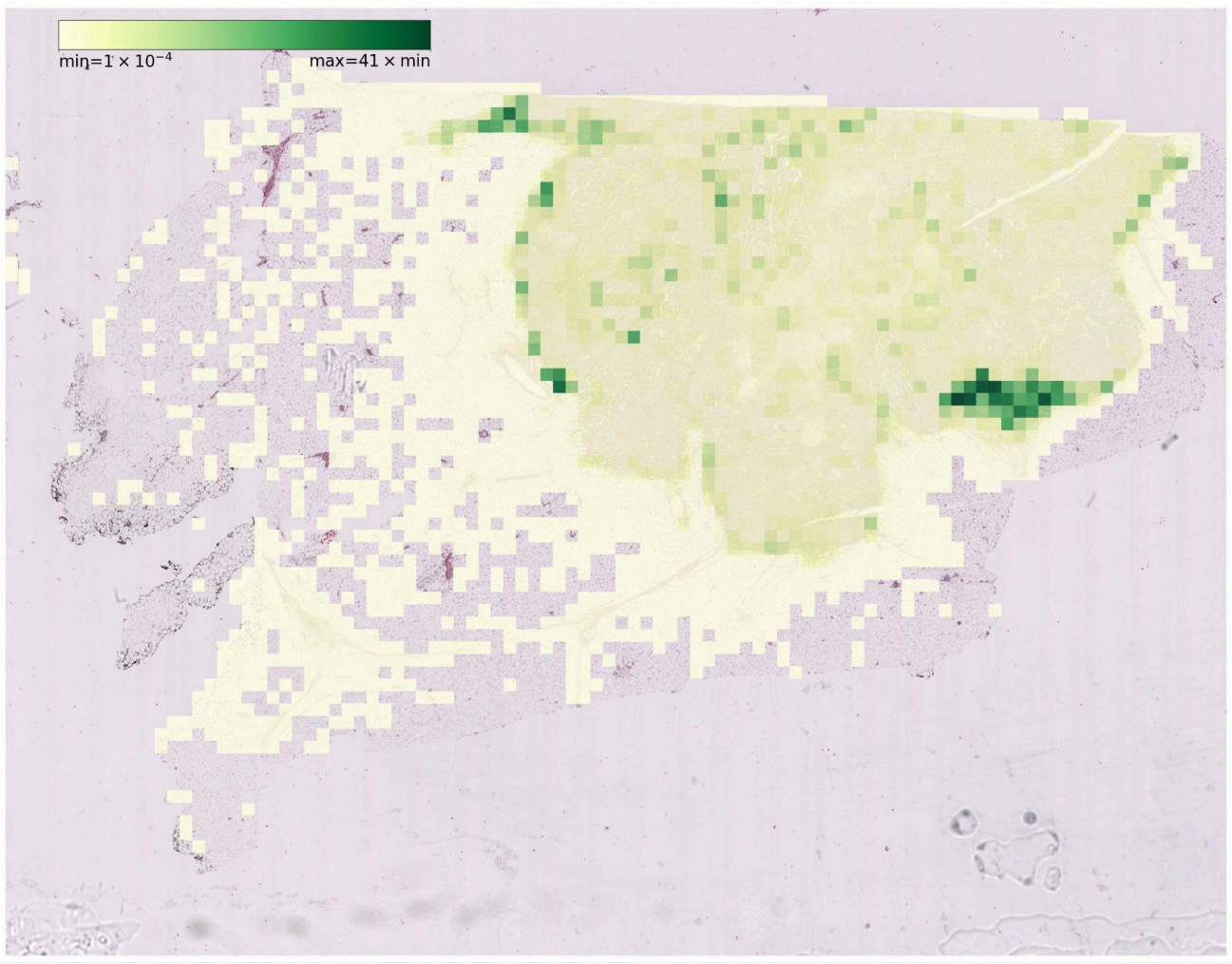

d. Score heatmap

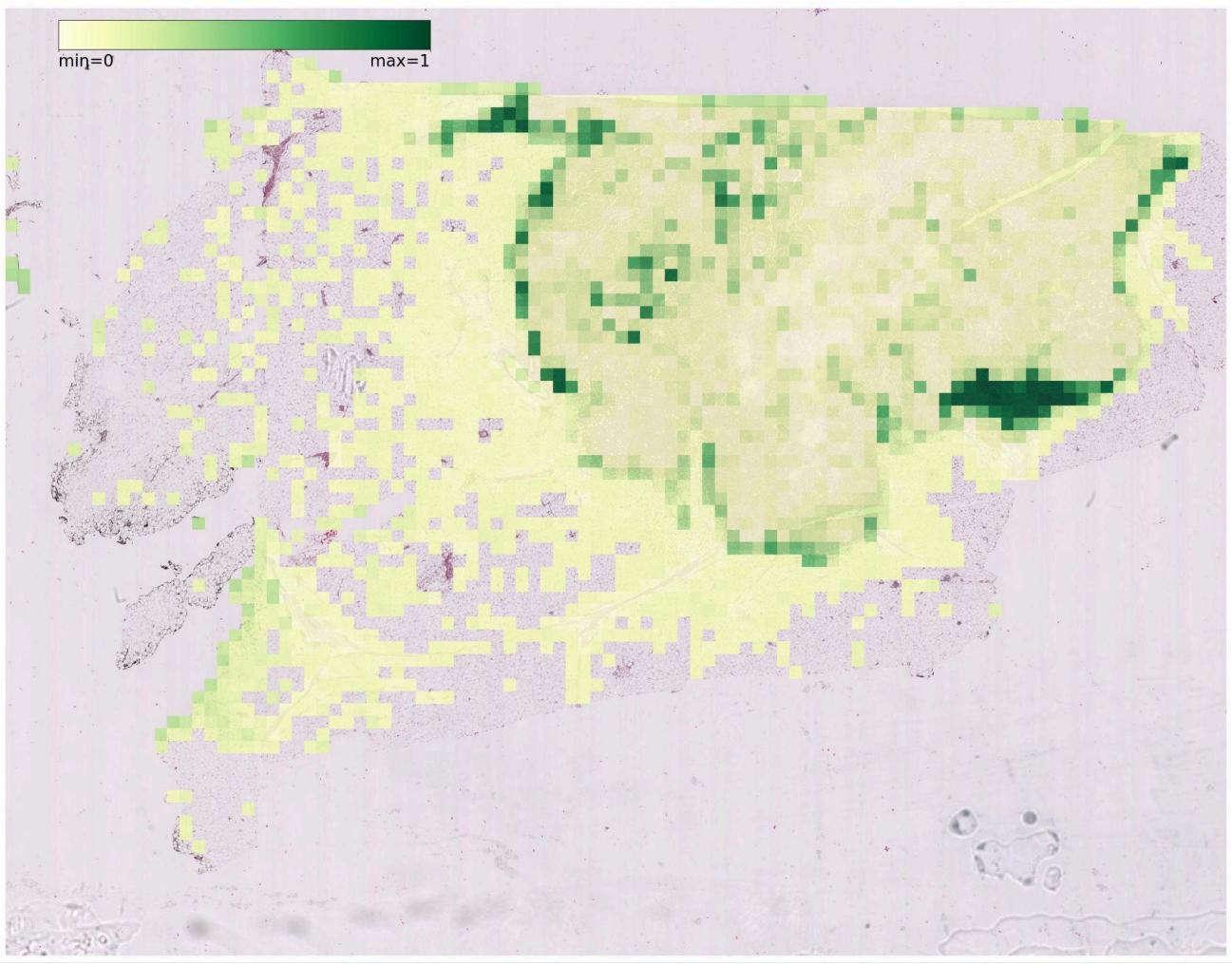

Sample from BASIS with a high pathologist score (83%) and a low ECTIL score (30%).

The slide quality and tissue mask are sufficient, and the provided pathologist's TILs score is correct. It is evident that there is a large fibrotic area, and a large area of fatty tissue included in the model. The scores provided within the tumour bed appear correct. However, due to the inclusion of necrotic area and fatty tissue in the model, the cumulative weight of these tiles that are not of interest dilute the score, resulting in severe underprediction.

**Figure S5-2: Discordant case of BASIS**

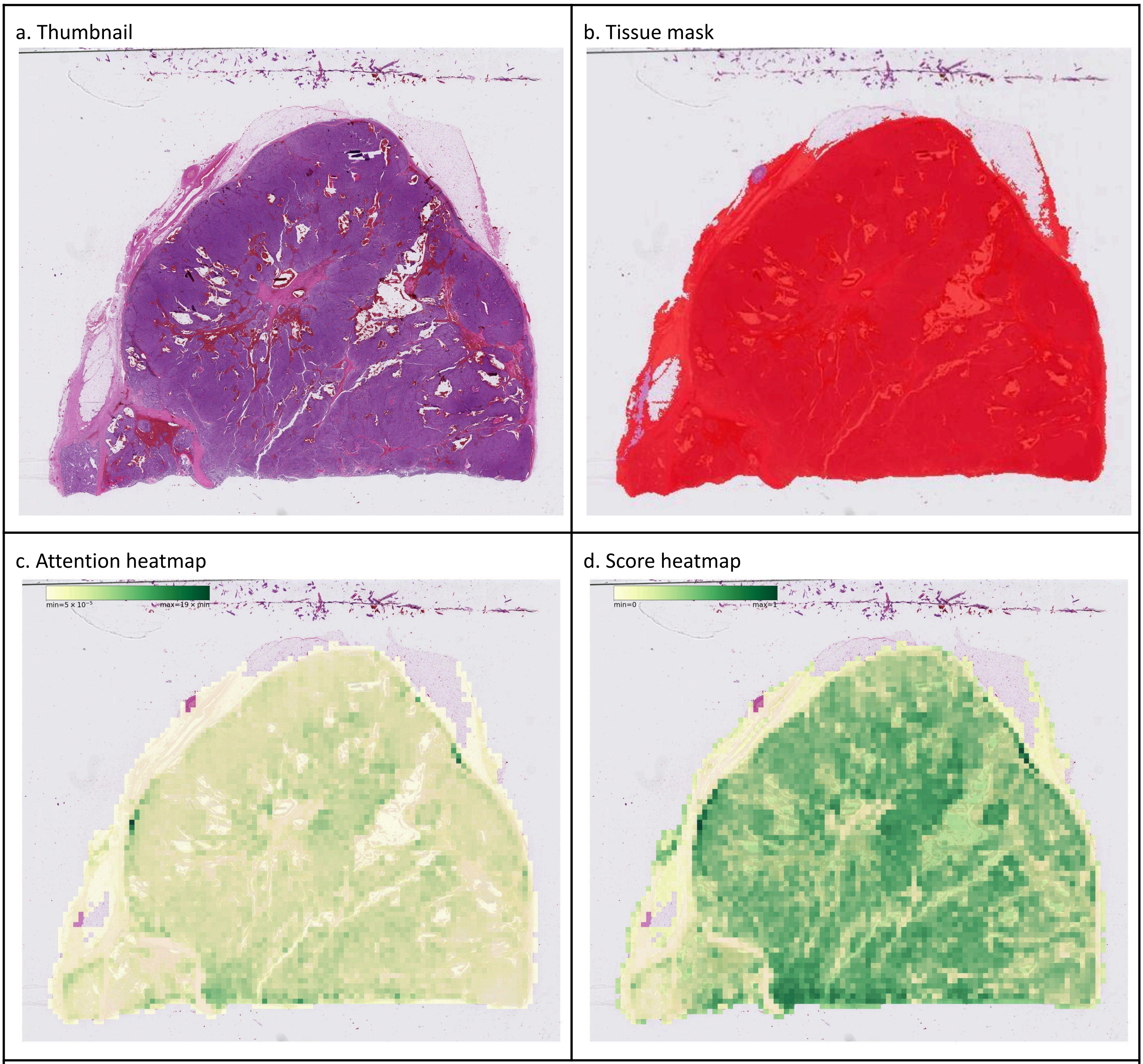

Sample from BASIS with a low pathologist score (8%) and a high ECTIL score (41%). The pathologist noted that the slide quality is high and the tissue mask looks correct. However, there's a very strong focus on the tumour border with a very high attention weight and TILs score. A cumulative attention weight analysis (not shown) displays that the majority of the attention weight comes from these areas, which leads to an overprediction. Since the tumour border contains many TILs outside of the tumour, these are not included according to the TILS-WG guidelines. The reason for the overprediction is due to a lack of context; were the model to recognize that these TILs lie outside the tumour, they may have been properly excluded. The rest of the tissue is scored well by ECTIL.

**Figure S6-1: Discordant case of N4+**

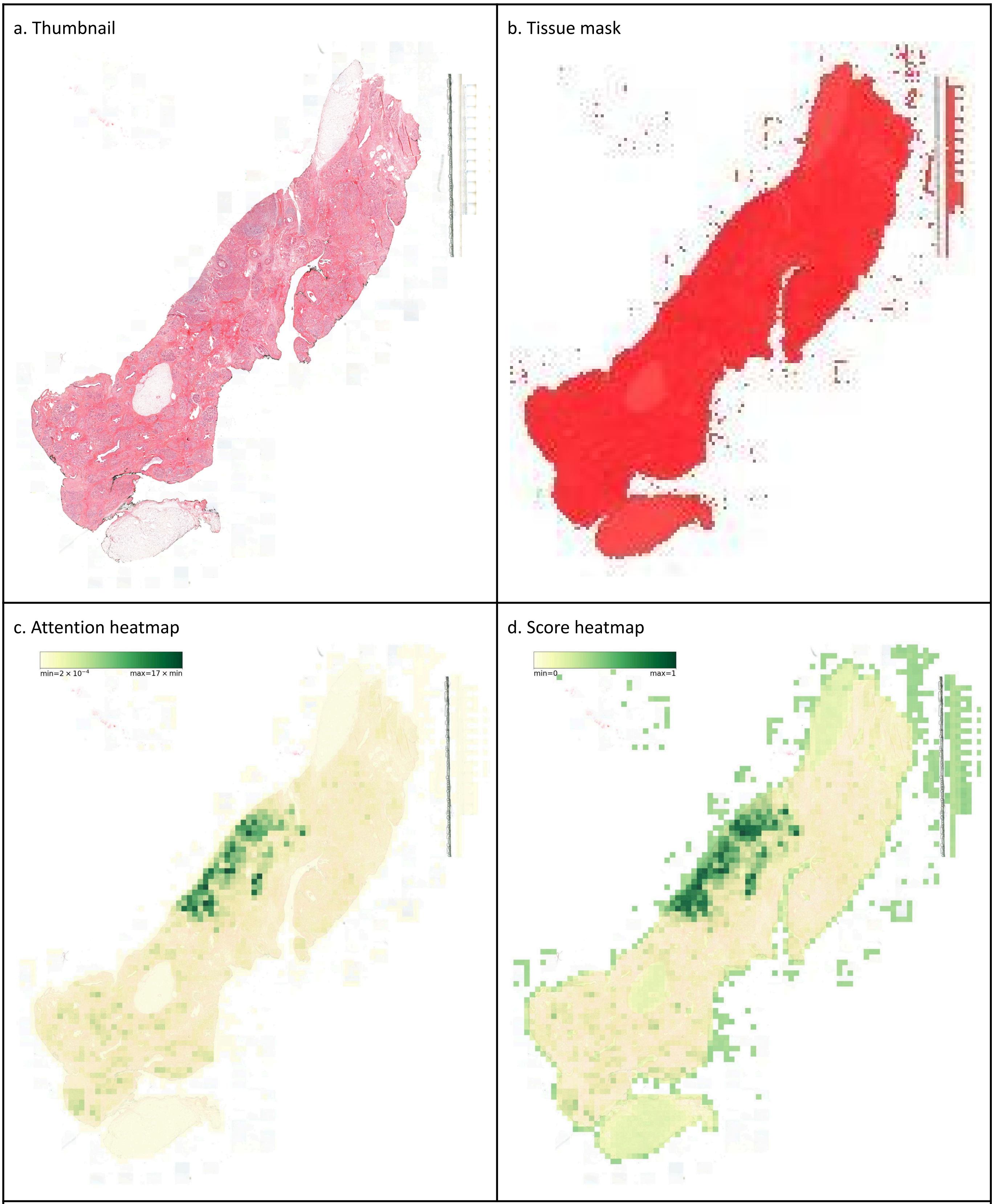

Sample from N4+ with a high pathologist score (80%) and a low ECTIL score (27%).

The pathologist noted that the slide quality is low, looking pale with crushed artefacts and blurry regions. The tissue mask looks good. In general, the heatmaps appear to focus on meaningful regions. The tumour bed is very small, and the area of lymphovascular invasion is larger than the tumour bed. While the model includes the tumour bed with TILs and lymphovascular invasion with TILs, this only covers 10% of the slide. Hence, there are two limitations of ECTIL. First, it includes lymphovascular invasion, which should not be included, although this is not the source of the underprediction since it contains approximately as many TILs as the tumour area. More importantly, the score of the relevant regions is diluted through the large cumulative attention weights of the remaining 90% of the whole slide image that does not contain TILs.

**Figure S6-2: Discordant case of N4+**

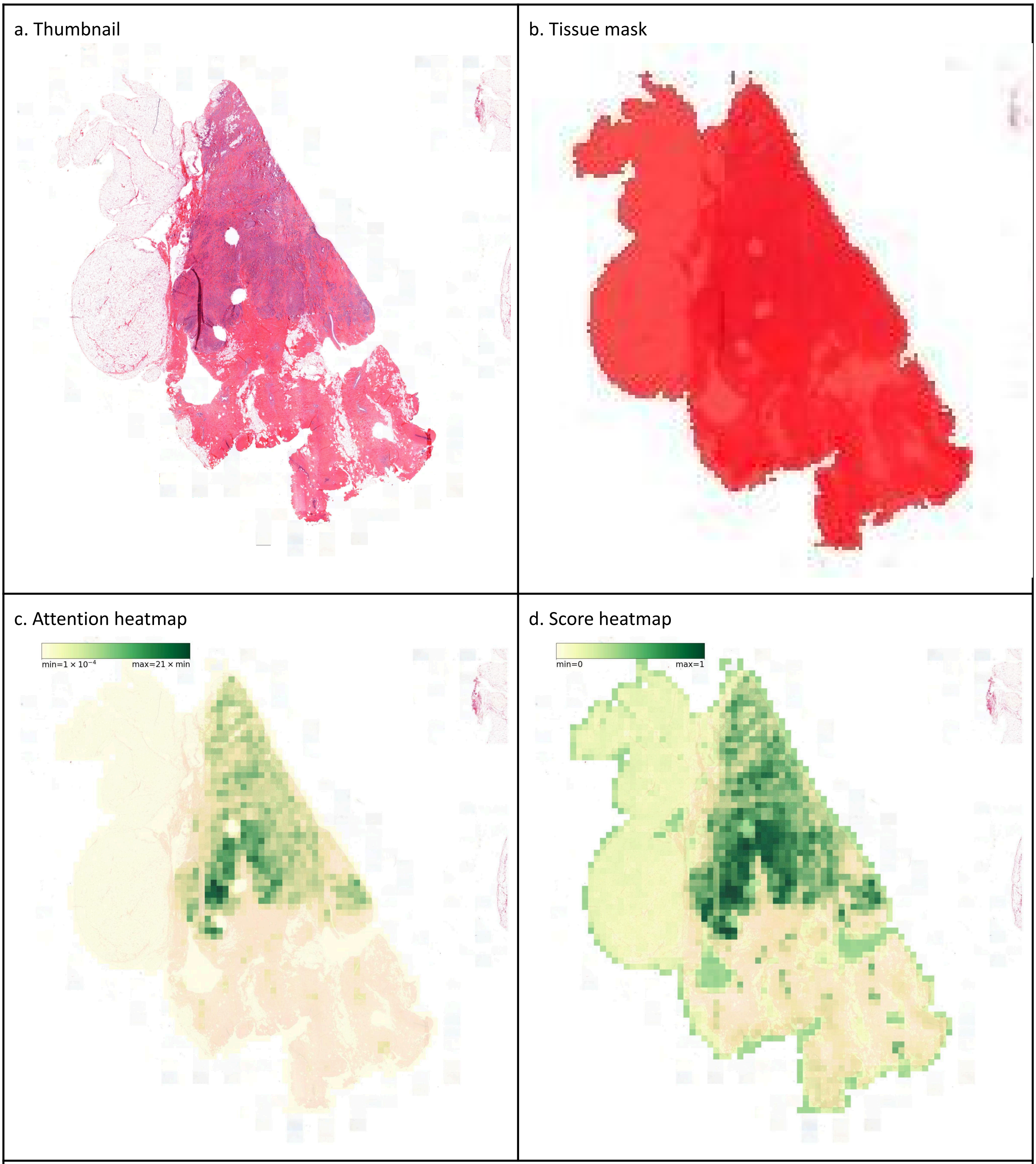

Sample from N4+ with a low pathologist score (10%) and a high ECTIL score (43%).

The slide quality and tissue mask are fine. This is a case of invasive lobular carcinoma, receiving very high TILs scores in the tumour bed, due to the morphological similarity between plasma cells, lymphocytes, and lobular tumour cells. The remaining fatty and stromal areas dilute the score down to 43%, which remains too high a score.

**Figure S7-1: Discordant case of MATADOR**

| a. Thumbnail | b. Tissue mask |
|---|---|

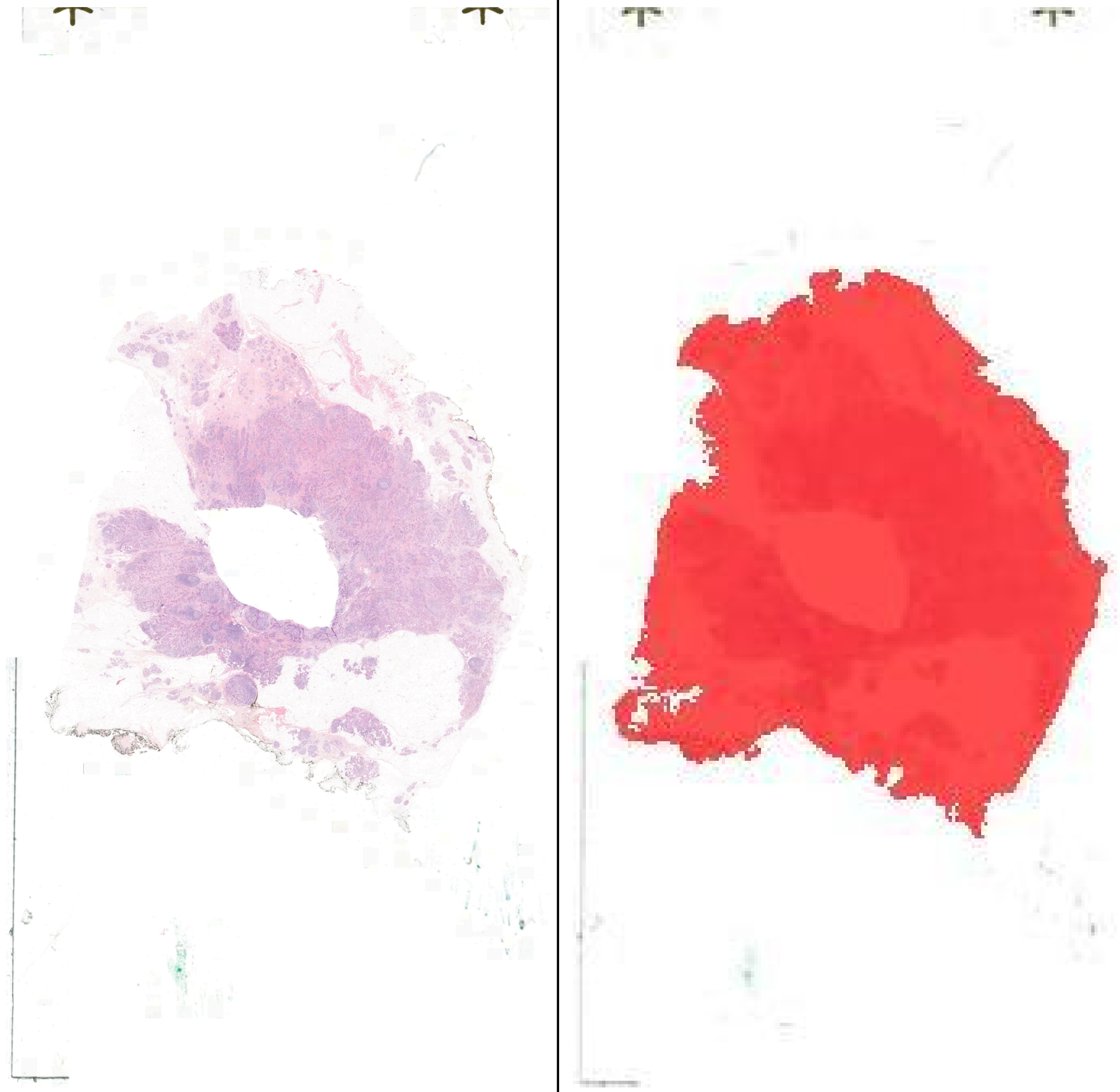

| c. Attention heatmap | d. Score heatmap |
|---|---|

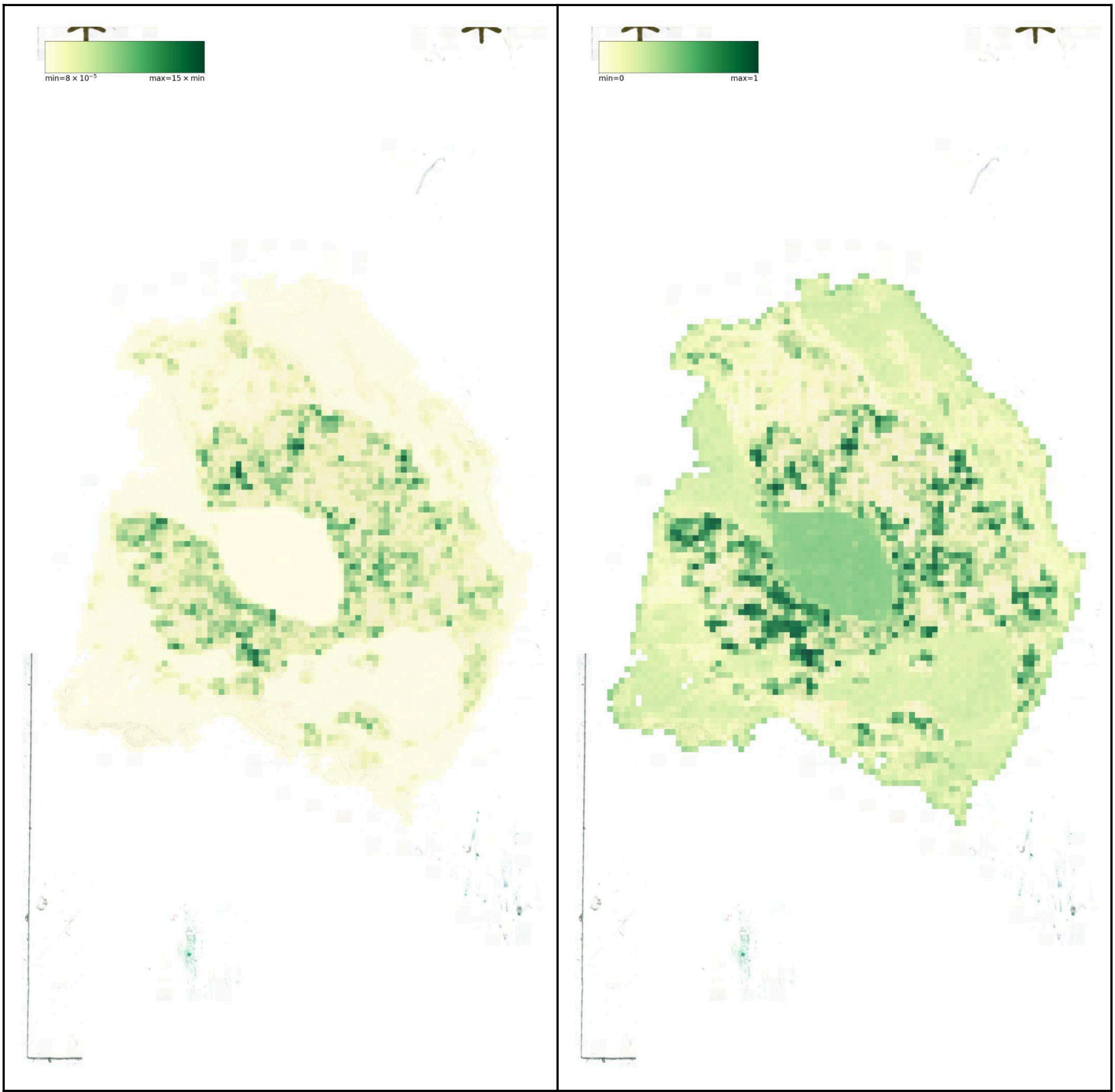

Sample from MATADOR with a high pathologist score (55%) and a low ECTIL score (34%).

Pathologist finds that the slide quality and tissue mask are correct. It is immediately evident there are many lymphocytic hotspots which should not be included. The existing pathologist score (55%) appears to be too high, likely due to confusion by the lymphocytic hotspots during the original scoring. The model is evidently providing high attention and TILs scores to these lymphocytic hotspots, since the lack of context makes the model unaware of whether it is a lymphocytic hotspot or a densely lymphocyte-populated area within the tumour bed. Consequently, these very high scores are diluted due to the large non-tumoural area included. However, with revision the pathologist score would likely be lowered, bringing the model and pathologist score in close proximity.

**Figure S7-2: Discordant case of MATADOR**

| | |
|---|---|
| a. Thumbnail<br>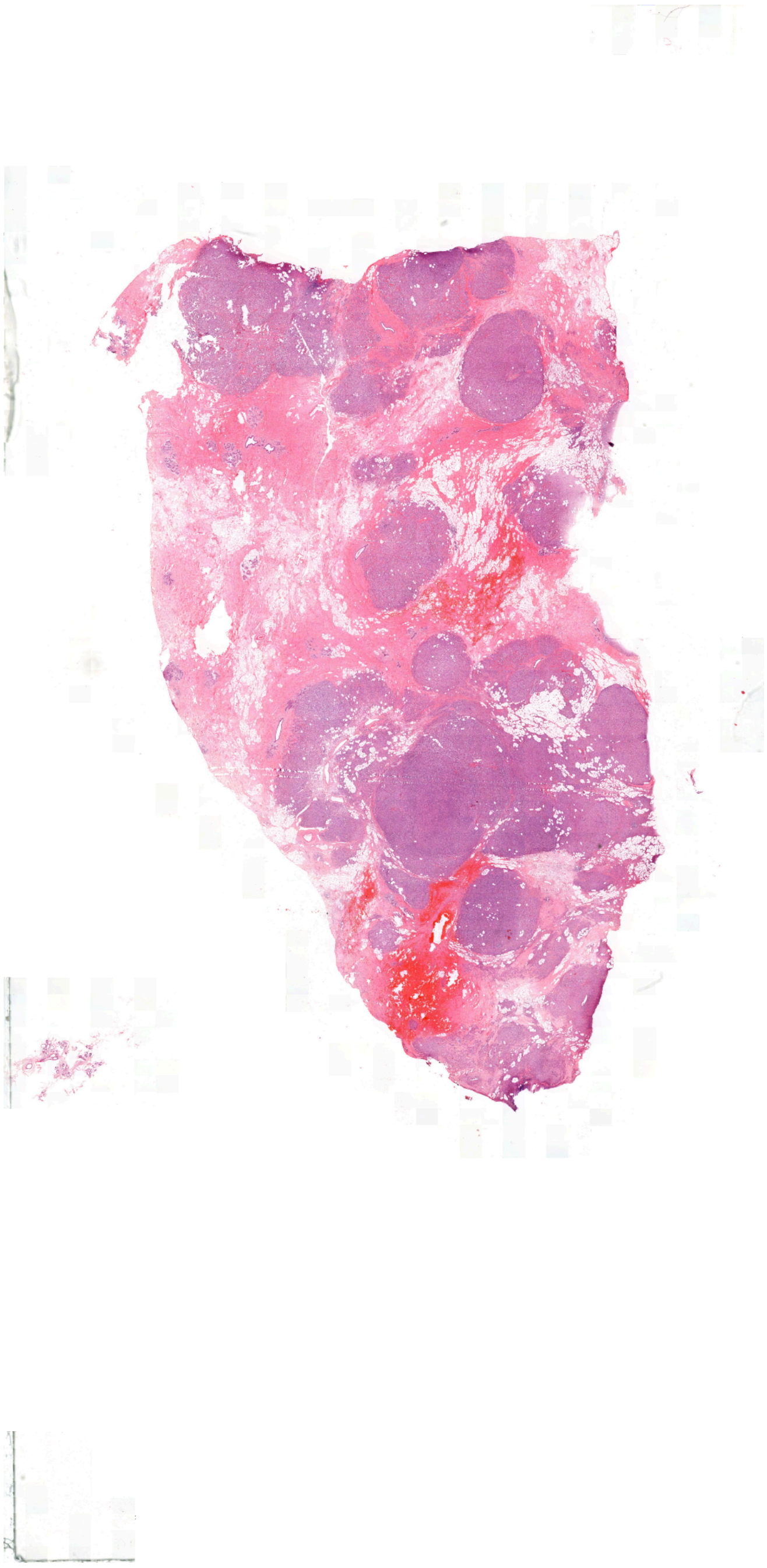 | b. Tissue mask<br>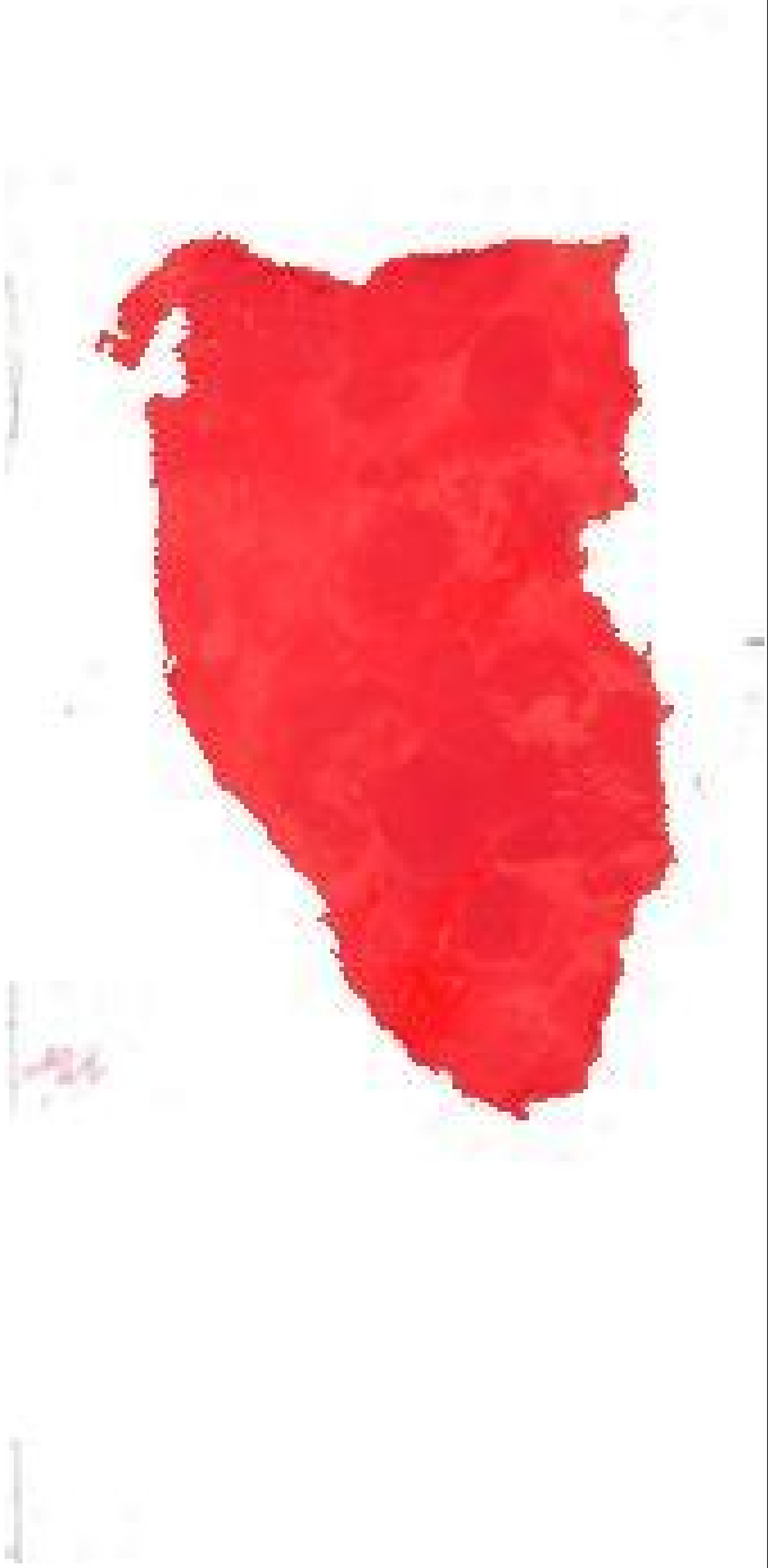 |
| c. Attention heatmap | d. Score heatmap |

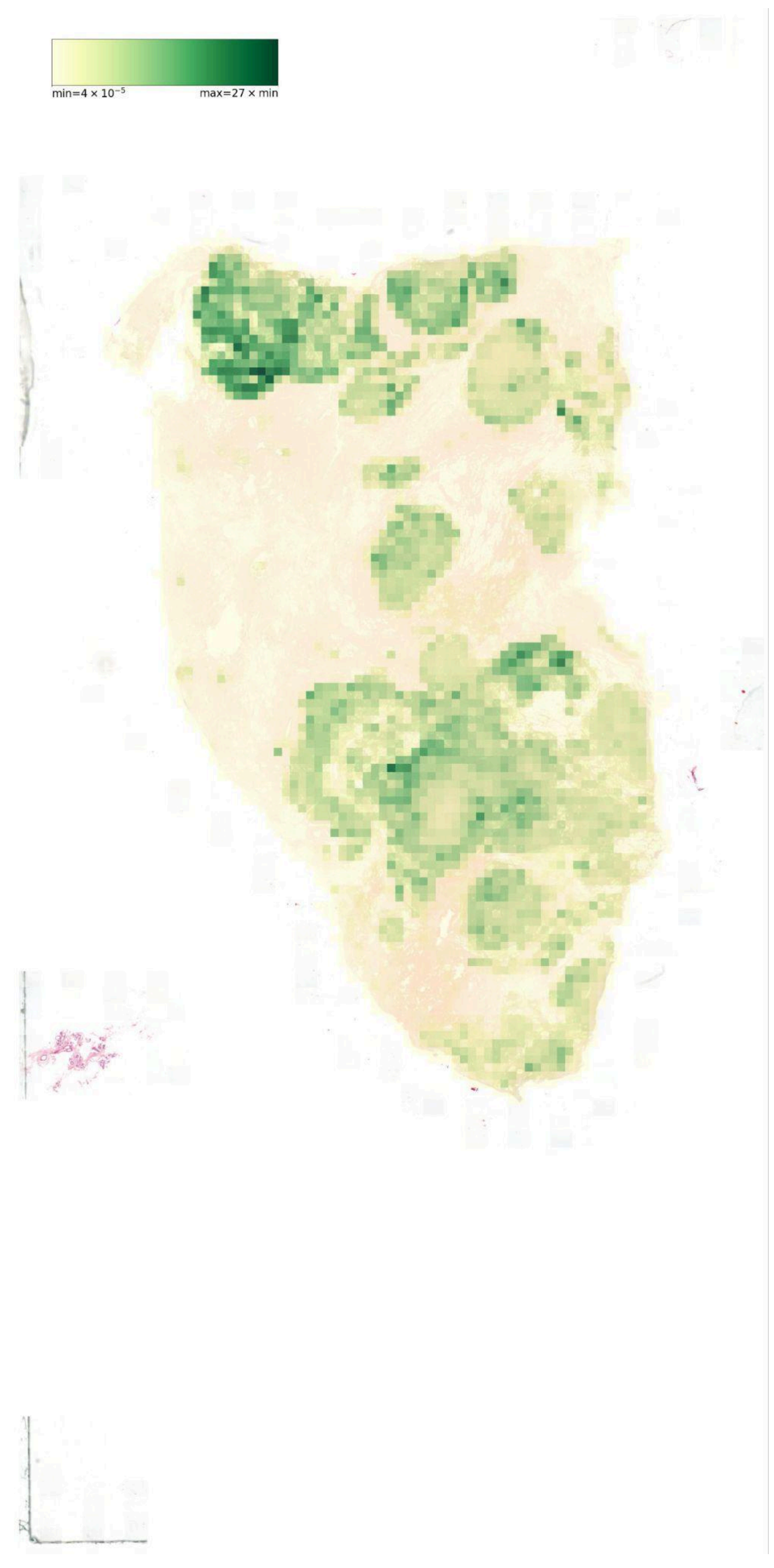

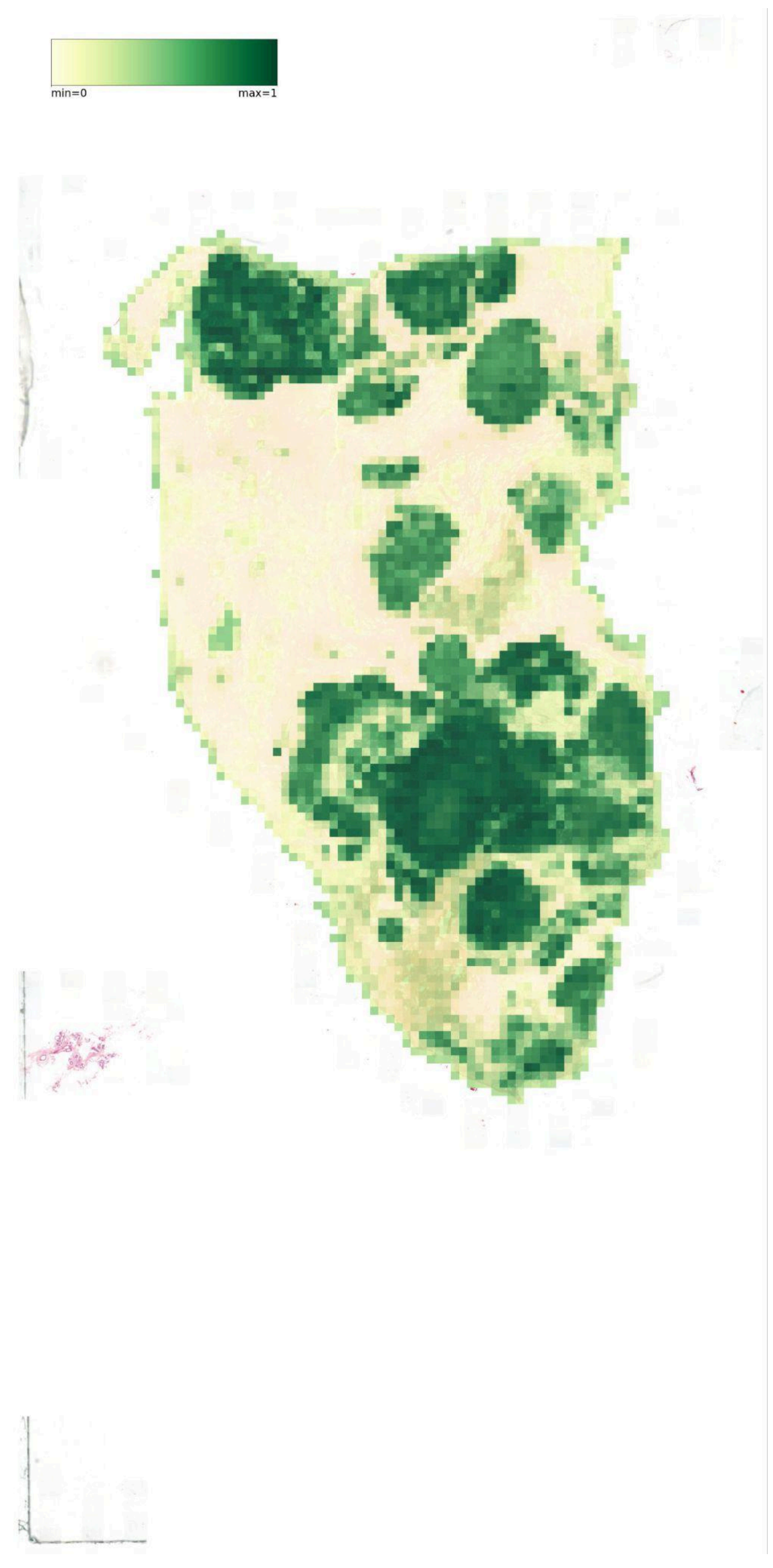

Sample from MATADOR with a low pathologist score (1%) and a high ECTIL score (55%).

Pathologist finds that the slide quality is high and the tissue mask is correct. The pathologist TILs score is 1%, but this is noted to be a difficult case. Since this is a case of invasive lobular carcinoma, the tumour cells look like lymphocytes, which are scored as such. Our model scores these to be TILs and ends up with a very high TILs score.

**Figure S8: Survival curves based on pathologist's TILs scores and previously defined cutoffs**

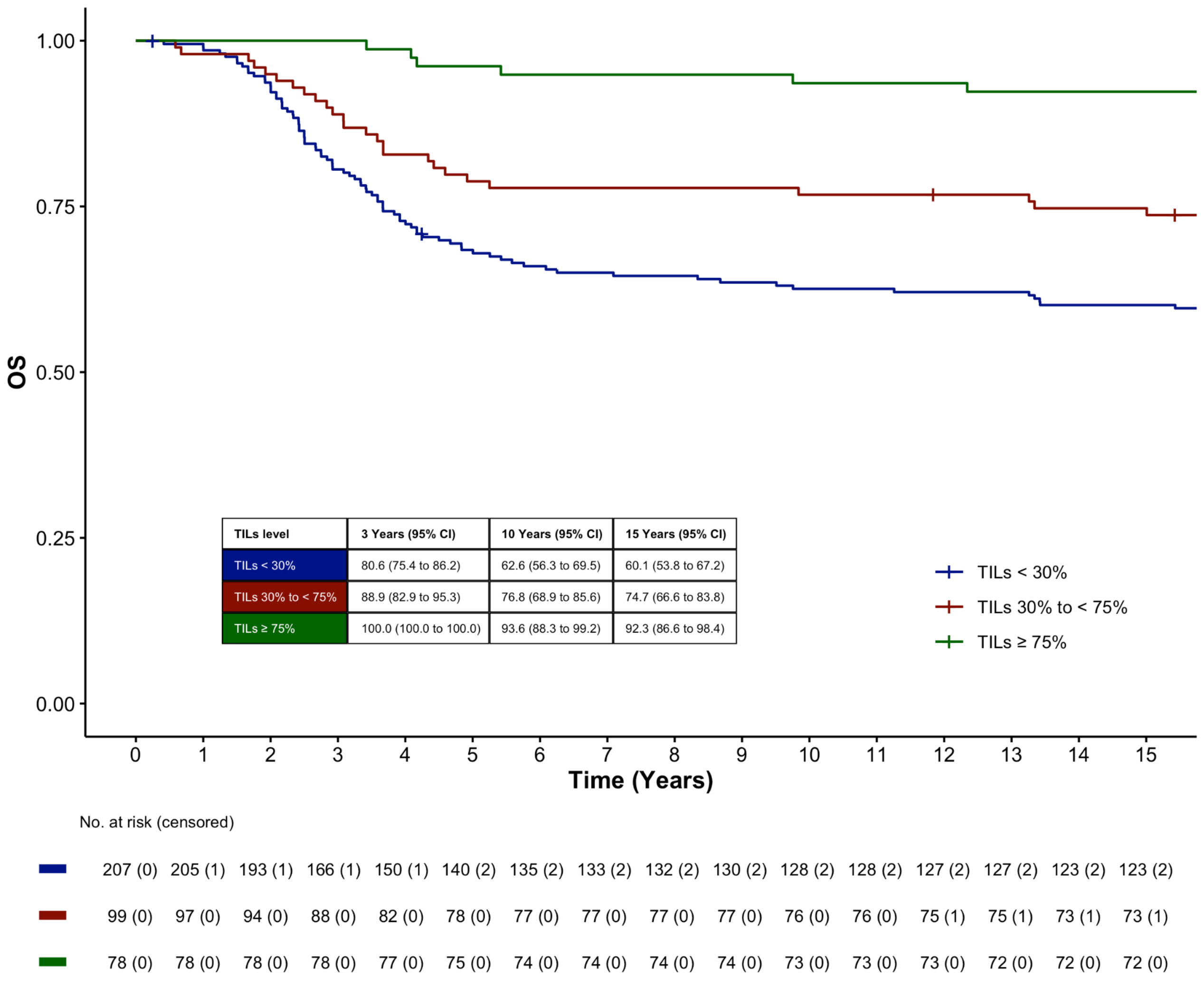

Kaplan-meier plot of overall survival using the pathologist scores on PARADIGM with the same cutoffs as used in de Jong et al. ($p<0{\cdot}001$)

**Figure S9: Survival curves based on pathologist's TILs scores and median score cutoff**

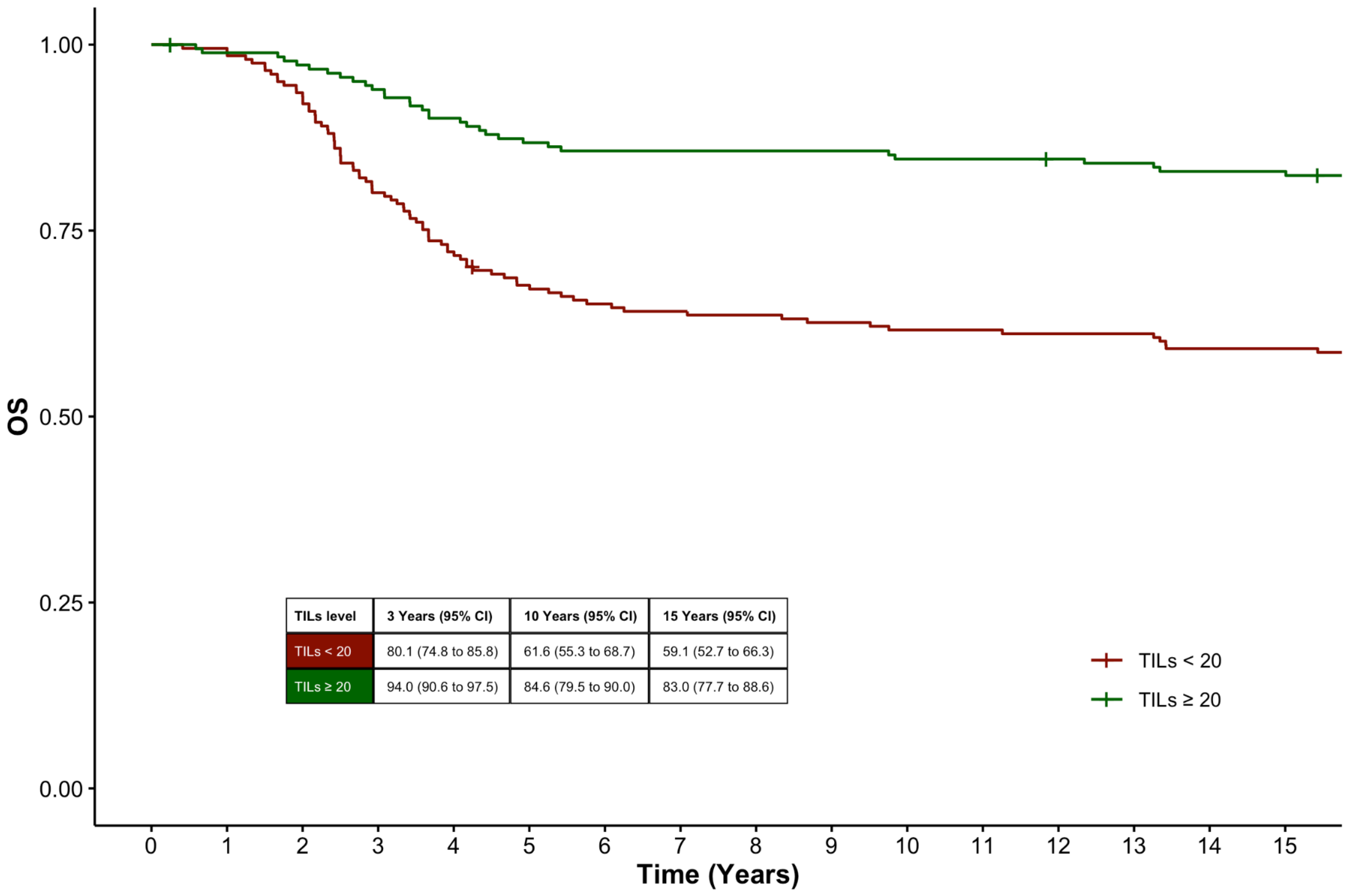

| TILs level | 3 Years (95% CI) | 10 Years (95% CI) | 15 Years (95% CI) |
|---|---|---|---|
| TILs < 20 | 80.1 (74.8 to 85.8) | 61.6 (55.3 to 68.7) | 59.1 (52.7 to 66.3) |
| TILs ≥ 20 | 94.0 (90.6 to 97.5) | 84.6 (79.5 to 90.0) | 83.0 (77.7 to 88.6) |

No. at risk (censored)

| | 0 | 1 | 2 | 3 | 4 | 5 | 6 | 7 | 8 | 9 | 10 | 11 | 12 | 13 | 14 | 15 |
|---|---|---|---|---|---|---|---|---|---|---|---|---|---|---|---|---|
| TILs < 20 | 201 (0) | 200 (0) | 188 (0) | 161 (0) | 145 (0) | 135 (1) | 130 (1) | 128 (1) | 127 (1) | 125 (1) | 123 (1) | 123 (1) | 122 (1) | 122 (1) | 118 (1) | 118 (1) |
| TILs ≥ 20 | 183 (0) | 180 (1) | 177 (1) | 171 (1) | 164 (1) | 158 (1) | 156 (1) | 156 (1) | 156 (1) | 156 (1) | 154 (1) | 154 (1) | 153 (2) | 152 (2) | 150 (2) | 150 (2) |

Kaplan-meier plot of overall survival using the pathologist's scores on PARADIGM, split by the median score of 20 (p<0·001)

**Figure S10: Survival curves based on ECTIL-combined TILs scores and the median score cutoff**

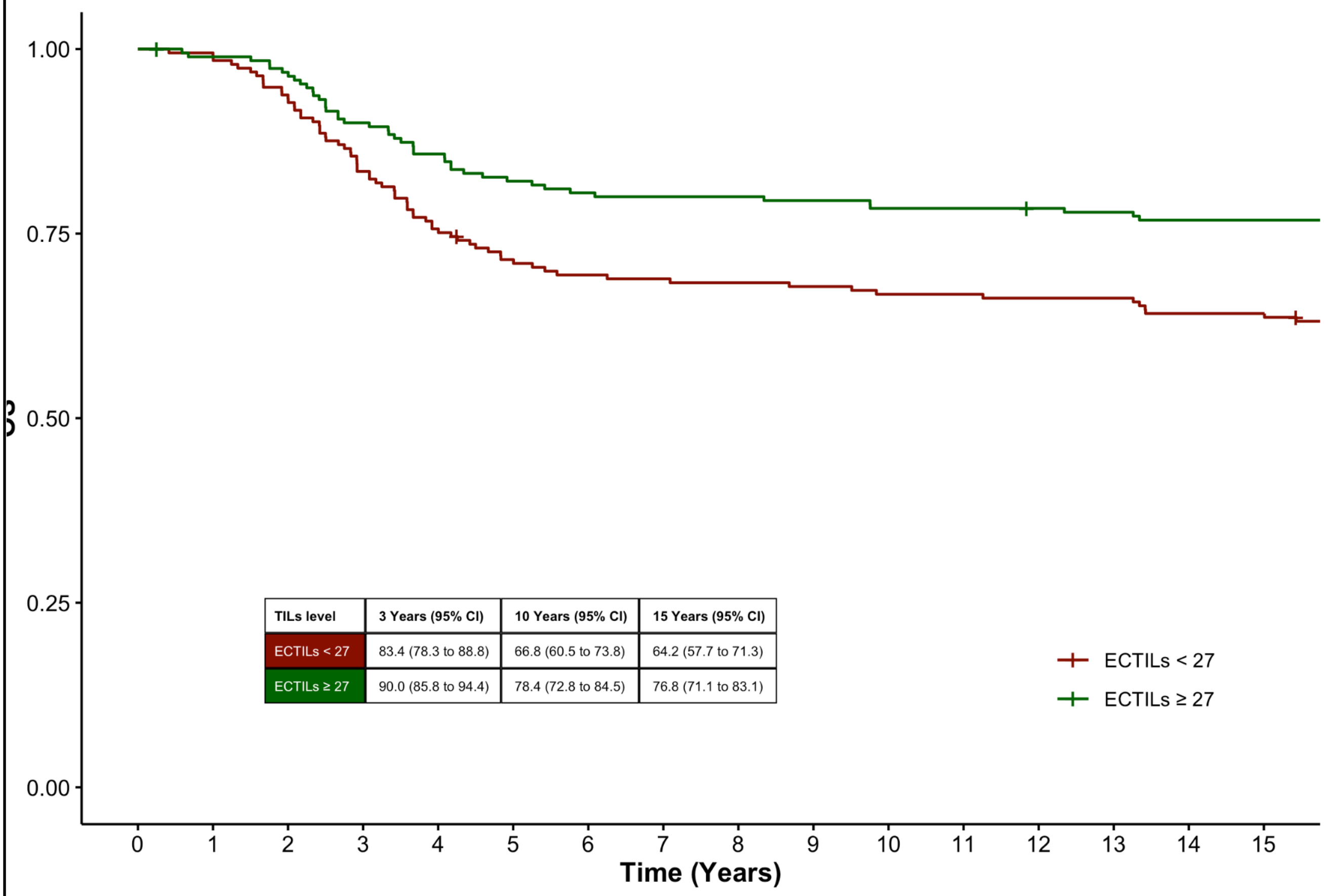

No. at risk (censored)

| | 0 | 1 | 2 | 3 | 4 | 5 | 6 | 7 | 8 | 9 | 10 | 11 | 12 | 13 | 14 | 15 |
|---|---|---|---|---|---|---|---|---|---|---|---|---|---|---|---|---|
| ECTILs < 27 | 193 (0) | 192 (0) | 181 (0) | 161 (0) | 146 (0) | 137 (1) | 133 (1) | 132 (1) | 131 (1) | 130 (1) | 128 (1) | 128 (1) | 127 (1) | 127 (1) | 123 (1) | 123 (1) |
| ECTILs ≥ 27 | 191 (0) | 188 (1) | 184 (1) | 171 (1) | 163 (1) | 156 (1) | 153 (1) | 152 (1) | 152 (1) | 151 (1) | 149 (1) | 149 (1) | 148 (2) | 147 (2) | 145 (2) | 145 (2) |

Kaplan-Meier plot of overall survival using the ECTIL scores on PARADIGM, split by the median score of 27 ($p<0{\cdot}01$)

**Supplementary tables**

**Table S1: Patient characteristics of combined cohorts**

| | ECTIL-TNBC cohort (N=400) | ECTIL-combined cohort** (N=1964) |
|---|---|---|
| **Age** | | |
| Median [Min, Max] | 48·0 [26·0, 90·0] | 49·0 [22·0, 90·0] |
| Missing | 0 (0%) | 14 (0·7%)** |
| **Gender** | | |
| Female | 400 (100%) | 1943 (98·9%) |
| Male | 0 (0%) | 7 (0·4%) |
| Missing | 0 (0%) | 14 (0·7%)** |
| **Menopausal status** | | |
| Premenopausal | 61 (15·3%) | 484 (24·6%) |
| Perimenopausal | 1 (0·3%) | 24 (1·2%) |
| Postmenopausal | 60 (15·0%) | 483 (24·6%) |
| Missing | 278 (69·5%) | 973 (49·5%)** |
| **Disease stage** | | |
| I | 39 (9·8%) | 140 (7·1%) |
| II | 153 (38·3%) | 842 (42·9%) |
| III | 175 (43·8%) | 864 (44·0%) |
| IV | 1 (0·3%) | 12 (0·6%) |
| Missing | 32 (8·0%) | 106 (5·4%)** |
| **T-stage** | | |
| T0 | 0 (0%) | 1 (0·1%)* |
| T1 | 84 (21·0%) | 441 (22·5%) |
| T2 | 148 (37·0%) | 855 (43·5%) |
| T3 | 25 (6.3%) | 209 (10·6%) |
| T4 | 1 (0·3%) | 22 (1·1%) |
| Tx | 27 (6.8%) | 69 (3·5%) |
| Missing | 115 (28·8%) | 367 (18·7%)** |
| **N-stage** | | |
| N0 | 46 (11·5%) | 304 (15·5%) |
| N1 | 58 (14·5%) | 511 (26·0%) |
| N2 | 11 (2·8%) | 120 (6·1%) |
| N3 | 2 (0·5%) | 44 (2·2%) |
| NX | 29 (7·3%) | 77 (3·9%) |
| Missing | 254 (63·5%) | 908 (46·2%)** |
| **M-stage** | | |
| M0 | 244 (61·0%) | 1432 (72·9%) |
| MX | 44 (11·0%) | 174 (8·9%) |
| M1 | 0 (0%) | 2 (0·1%) |
| Missing | 112 (28·0%) | 356 (18·1%)** |
| **Tumour grade** | | |
| 1 | 9 (2·3%) | 218 (11·1%) |
| 2 | 38 (9·5%) | 655 (33·4%) |
| 3 | 230 (57·5%) | 675 (34·4%) |
| Missing | 123 (30·8%) | 416 (21·2%)** |
| **Lymphovascular invasion** | | |
| Absent | 42 (10·5%) | 182 (9·3%) |
| Present | 7 (1·8%) | 40 (2·0%) |
| Missing | 351 (87·8%) | 1742 (88·7%)** |
| **Breast cancer subtype** | | |
| ER+/HER2+ | 0 (0%) | 230 (11·7%) |
| ER+/HER2- | 0 (0%) | 1175 (59·8%) |
| ER-/HER2+ | 0 (0%) | 123 (6·3%) |
| TN | 400 (100%) | 400 (20·4%) |
| Missing | 0 (0%) | 36 (1·8%)** |
| **Histology** | | |
| BC NST | 273 (68·3%) | 1308 (66·6%) |
| ILC | 15 (3·8%) | 169 (8·6%) |
| Metaplastic carcinoma | 2 (0·5%) | 3 (0·2%) |
| Other | 22 (5·5%) | 115 (5·9%) |
| Missing | 88 (22·0%) | 369 (18·8%)** |
| **TILs score pathologist (%)** | | |
| Median [Min, Max] | 25·0 [0·500, 93·0] | 10·0 [0, 93·0] |
| Missing | 0 (0%) | 14 (0·7%)** |

Table S1: Patient characteristics of combined cohorts ECTIL-TNBC and ECTIL-combined. ER: Estrogen receptor; HER2: Human Epidermal growth factor Receptor 2, TN: Triple Negative; BC NST: Breast Cancer of No Special Type; ILC: Invasive Lobular Carcinoma; TILs: Tumour Infiltrating Lymphocytes.

*Patient with no evidence of tumour in the breast but cancer has spread to axillary lymph nodes.

** For TCGA, 14 patients have two available slides, which are included in the sample count. In the table, we present patient characteristics; hence, for 14 samples, the clinical data is presented as missing..

**Table S2: Metrics of ECTIL-TCGA on all cohorts**

| | TCGA | BASIS | N4+ | TRAIN II | MATADOR | PARADIGM |
|---|---|---|---|---|---|---|
| Pearson r | 0·61 | 0·65 | 0·54 | 0·60 | 0·74 | 0·58 |
| Spearman r | 0·63 | 0·51 | 0·47 | 0·53 | 0·63 | 0·60 |
| concordance r | 0·55 | 0·54 | 0·49 | 0·28 | 0·51 | 0·38 |
| mse | 238·00 | 264·00 | 241·00 | 765·00 | 213·00 | 769·00 |
| AUROC@10 | 0·82 | 0·73 | 0·72 | 0·77 | 0·87 | 0·80 |
| AUROC@30 | 0·84 | 0·86 | 0·82 | 0·85 | 0·94 | 0·80 |
| AUROC@50 | 0·82 | 0·90 | 0·85 | 0·84 | 0·94 | 0·79 |
| AUROC@75 | 0·89 | 0·92 | 0·89 | 0·89 | 0·99 | 0·84 |
| random AP@10 | 0·51 | 0·56 | 0·50 | 0·35 | 0·37 | 0·60 |
| random AP@30 | 0·19 | 0·25 | 0·17 | 0·20 | 0·09 | 0·43 |
| random AP@50 | 0·08 | 0·13 | 0·08 | 0·14 | 0·04 | 0·38 |
| random AP@75 | 0·03 | 0·02 | 0·02 | 0·04 | 0·00 | 0·15 |
| AP@10 | 0·83 | 0·78 | 0·72 | 0·69 | 0·81 | 0·84 |
| AP@30 | 0·57 | 0·67 | 0·48 | 0·67 | 0·66 | 0·75 |
| AP@50 | 0·31 | 0·64 | 0·43 | 0·58 | 0·49 | 0·69 |
| AP@75 | 0·18 | 0·16 | 0·30 | 0·26 | 0·13 | 0·50 |

Table S2: Concordance results of ECTIL-TCGA. Here, the performance of TCGA is computed once by combining all test folds into a single dataset. Remaining studies report the metric for average of the prediction of the five models from five-fold validation. Area under the Receiver Operating Characteristic curve (AUROC) and Average Precision (AP) are computed for binarized ground truth scores at various cutoffs (reported as AUROC@cutoff and AP@cutoff, where the labels are greater than cutoff, or smaller than or equal to cutoff) The random AP is the label balance, and is the value that a random classifier would achieve. The mean squared error (mse) is computed for the TILs score represented as a percentage (0-100). We find that the concordance on external cohorts is not lower than those of the internal test set, meaning the method generalises well to varying unseen distributions.

**Table S3: Detailed metrics of ECTIL-combined and ECTIL-TNBC on PARADIGM**

| | ECTIL-combined | ECTIL-TNBC |
|---|---|---|
| Pearson r | 0·69 | 0·64 |
| Spearman r | 0·72 | 0·66 |
| concordance r | 0·47 | 0·39 |
| mse | 702·00 | 739·00 |
| AUROC@10 | 0·86 | 0·82 |
| AUROC@30 | 0·85 | 0·83 |
| AUROC@50 | 0·85 | 0·82 |
| AUROC@75 | 0·9 | 0·88 |
| AP@10 (random=0·6) | 0·89 | 0·86 |
| AP@30 (random=0·43) | 0·82 | 0·78 |
| AP@50 (random=0·38) | 0·78 | 0·73 |
| AP@75 (random=0·15) | 0·62 | 0·57 |

Concordance results of ECTIL-TNBC and ECTIL-combined. Area under the Receiver Operating Characteristic curve (AUROC) and Average Precision (AP) are computed for binarized ground truth scores at various cutoffs (reported as AUROC@cutoff and AP@cutoff, where the labels are greater than cutoff, or smaller than or equal to cutoff) The random AP is the label balance, and is the value that a random classifier would achieve.

**Table S4: Subset analysis**

| Study | Clinical variable | Variable categories | Pearson's r | AUROC | n |
|---|---|---|---|---|---|
| BASIS | Menopausal status | Perimenopausal | 0·96 | 1 | 4 |
| | | Postmenopausal | 0·52 | 0·76 | 128 |
| | | Premenopausal | 0·72 | 0·89 | 58 |
| | T-stage | T1 | 0·79 | 0·88 | 59 |
| | | T2 | 0·52 | 0·76 | 92 |
| | | T3 | 0·13 | 0·84 | 20 |
| | | T4 | 0·43 | 1 | 8 |
| | | Tx | 0·7 | 0·82 | 68 |
| | N-stage | N0 | 0·59 | 0·79 | 83 |
| | | N1 | 0·65 | 0·82 | 58 |
| | | N2 | 0·36 | 0·82 | 25 |
| | | N3 | 0·35 | 0·67 | 4 |
| | | NX | 0·7 | 0·84 | 77 |
| | M-stage | M0 | 0·68 | 0·86 | 73 |
| | | M1 | 1 | 1 | 2 |
| | | MX | 0·63 | 0·8 | 172 |
| | Breast cancer subtype | ER+/HER2+ | 0·75 | 0 | 3 |
| | | ER+/HER2- | 0·5 | 0·67 | 188 |
| | | ER-/HER2+ | -1 | 0 | 2 |
| | | TN | 0·67 | 0·84 | 54 |
| | Tumour grade | 1 | 0·33 | 0·52 | 32 |
| | | 2 | 0·53 | 0·64 | 90 |
| | | 3 | 0·68 | 0·86 | 124 |
| | Lymphovascular invasion | Absent | 0·64 | 0·81 | 182 |
| | | Present | 0·76 | 0·91 | 40 |
| | Histology | BC NST | 0·71 | 0·87 | 193 |
| | | ILC | 0·5 | 0·78 | 17 |
| | | Metaplastic carcinoma | -0·59 | 0 | 3 |
| | | Other | -0·05 | 0·37 | 34 |
| | Disease stage | I | 0·81 | 0·91 | 41 |
| | | II | 0·55 | 0·76 | 85 |
| | | III | 0·33 | 0·82 | 43 |
| MATADOR | Menopausal status | Postmenopausal | 0·79 | 0·86 | 245 |
| | | Premenopausal | 0·68 | 0·86 | 266 |
| | T-stage | T1 | 0·75 | 0·88 | 236 |
| | | T2 | 0·75 | 0·86 | 249 |
| | | T3 | 0·5 | 0·81 | 29 |
| | | T4 | 1 | 0 | 2 |
| | N-stage | N0 | 0·75 | 0·9 | 107 |
| | | N1 | 0·74 | 0·85 | 315 |
| | | N2 | 0·76 | 0·88 | 74 |
| | | N3 | 0·65 | 0·76 | 21 |
| | M-stage | M0 | 0·74 | 0·87 | 517 |
| | Breast cancer subtype | ER+/HER2+ | 0·74 | 0·79 | 8 |
| | | ER+/HER2- | 0·68 | 0·84 | 409 |
| | | ER-/HER2+ | 0·95 | 0·96 | 7 |
| | | TN | 0·72 | 0·82 | 92 |
| | Tumour grade | 1 | 0·73 | 0·81 | 53 |
| | | 2 | 0·64 | 0·77 | 220 |
| | | 3 | 0·75 | 0·88 | 221 |
| | Histology | BC NST | 0·74 | 0·87 | 414 |
| | | ILC | 0·51 | 0·77 | 71 |
| | | Other | 0·84 | 1 | 25 |
| | Disease stage | I | 0·71 | 0·88 | 55 |
| | | II | 0·75 | 0·86 | 352 |
| | | III | 0·73 | 0·86 | 109 |
| N4+ | T-stage | T1 | 0·41 | 0·73 | 129 |
| | | T2 | 0·6 | 0·79 | 328 |
| | | T3 | 0·3 | 0·55 | 85 |
| | M-stage | M0 | 0·54 | 0·76 | 552 |
| | Breast cancer subtype | ER+/HER2- | 0·48 | 0·72 | 410 |
| | | TN | 0·54 | 0·74 | 142 |
| | Tumour grade | 1 | 0·44 | 0·74 | 121 |
| | | 2 | 0·6 | 0·78 | 207 |
| | | 3 | 0·46 | 0·68 | 205 |
| | Histology | BC NST | 0·61 | 0·85 | 115 |
| | | ILC | 0·42 | 0·91 | 50 |
| | | Other | 0·52 | 0·77 | 39 |
| | Disease stage | III | 0·54 | 0·76 | 552 |
| PARADIGM | Menopausal status | Premenopausal | 0·58 | 0·79 | 390 |
| | T-stage | T1 | 0·63 | 0·84 | 232 |
| | | T2 | 0·51 | 0·74 | 151 |
| | | T3 | 0·49 | 0·75 | 6 |
| | N-stage | N0 | 0·58 | 0·79 | 390 |
| | M-stage | M0 | 0·58 | 0·79 | 390 |
| | Breast cancer subtype | TN | 0·58 | 0·79 | 390 |
| | Tumour grade | 1 | | 0 | 2 |
| | | 2 | 0·4 | 0·73 | 52 |
| | | 3 | 0·6 | 0·8 | 336 |
| | Lymphovascular invasion | Absent | 0·57 | 0·79 | 349 |
| | | Present | 0·7 | 0·83 | 40 |
| | Histology | BC NST | 0·59 | 0·79 | 333 |
| | | ILC | | 0 | 1 |
| | | Metaplastic carcinoma | 0·38 | 0·76 | 20 |
| | | Other | 0·55 | 0·79 | 36 |
| | Disease stage | I | 0·63 | 0·84 | 232 |
| | | II | 0·51 | 0·74 | 157 |
| TCGA | Breast cancer subtype | ER+/HER2+ | 0·5 | 0·75 | 32 |
| | | ER+/HER2- | 0·69 | 0·85 | 178 |
| | | ER-/HER2+ | 0·06 | 0 | 9 |
| | | TN | 0·49 | 0·76 | 116 |
| | Histology | BC NST | 0·62 | 0·84 | 335 |
| | | ILC | 0·27 | 0·83 | 18 |
| | | Other | 0·98 | 0 | 3 |
| | Disease stage | I | 0·65 | 0·84 | 45 |
| | | II | 0·62 | 0·83 | 214 |
| | | III | 0·51 | 0·81 | 73 |
| | | IV | 0·81 | 0 | 10 |
| TRAIN II | Menopausal status | Perimenopausal | 0·4 | 0·81 | 20 |
| | | Postmenopausal | 0·6 | 0·7 | 110 |
| | | Premenopausal | 0·61 | 0·77 | 160 |
| | T-stage | T0 | | 0 | 1 |
| | | T1 | 0·35 | 0·73 | 17 |
| | | T2 | 0·61 | 0·77 | 186 |
| | | T3 | 0·62 | 0·7 | 75 |
| | | T4 | 0·17 | 0·59 | 12 |
| | | Tx | | 0 | 1 |
| | N-stage | N0 | 0·56 | 0·7 | 114 |
| | | N1 | 0·66 | 0·75 | 138 |
| | | N2 | 0·34 | 0·85 | 21 |
| | | N3 | 0·66 | 0·76 | 19 |
| | M-stage | M0 | 0·6 | 0·74 | 290 |
| | | MX | -1 | 0 | 2 |
| | Breast cancer subtype | ER+/HER2+ | 0·65 | 0·73 | 187 |
| | | ER-/HER2+ | 0·53 | 0·75 | 105 |
| | Tumour grade | 1 | 0·76 | 0·55 | 12 |
| | | 2 | 0·54 | 0·69 | 138 |
| | | 3 | 0·61 | 0·77 | 125 |
| | Histology | BC NST | 0·62 | 0·76 | 265 |
| | | ILC | 0·19 | 0·68 | 13 |
| | | Other | 0·37 | 0·47 | 14 |
| | Disease stage | II | 0·62 | 0·74 | 201 |
| | | III | 0·56 | 0·74 | 91 |
| Aggregate | Menopausal status | Perimenopausal | 0·42 | 0·54 | 24 |
| | | Postmenopausal | 0·54 | 0·67 | 483 |
| | | Premenopausal | 0·6 | 0·76 | 874 |
| | T-stage | T0 | | 0 | 1 |
| | | T1 | 0·63 | 0·81 | 673 |
| | | T2 | 0·53 | 0·67 | 1006 |
| | | T3 | 0·39 | 0·56 | 215 |
| | | T4 | 0·14 | 0·5 | 22 |
| | | Tx | 0·69 | 0·82 | 69 |
| | N-stage | N0 | 0·57 | 0·75 | 694 |
| | | N1 | 0·6 | 0·67 | 511 |
| | | N2 | 0·49 | 0·75 | 120 |
| | | N3 | 0·56 | 0·73 | 44 |
| | | NX | 0·7 | 0·84 | 77 |
| | M-stage | M0 | 0·55 | 0·7 | 1822 |
| | | M1 | 1 | 1 | 2 |
| | | MX | 0·62 | 0·79 | 174 |
| | Breast cancer subtype | ER+/HER2+ | 0·46 | 0·58 | 230 |
| | | ER+/HER2- | 0·5 | 0·77 | 1185 |
| | | ER-/HER2+ | 0·4 | 0·66 | 123 |
| | | TN | 0·55 | 0·76 | 794 |
| | Tumour grade | 1 | 0·4 | 0·6 | 220 |
| | | 2 | 0·42 | 0·58 | 707 |
| | | 3 | 0·56 | 0·74 | 1011 |
| | Lymphovascular invasion | Absent | 0·6 | 0·78 | 531 |
| | | Present | 0·66 | 0·77 | 80 |
| | Histology | BC NST | 0·58 | 0·73 | 1655 |
| | | ILC | 0·36 | 0·64 | 170 |
| | | Metaplastic carcinoma | 0·37 | 0·69 | 23 |
| | | Other | 0·53 | 0·71 | 151 |
| | Disease stage | I | 0·66 | 0·83 | 373 |
| | | II | 0·55 | 0·7 | 1009 |
| | | III | 0·46 | 0·68 | 868 |
| | | IV | 0·81 | 0 | 10 |
| | Study | BASIS | 0·65 | 0·82 | 247 |
| | | MATADOR | 0·74 | 0·87 | 517 |
| | | N4+ | 0·54 | 0·76 | 552 |
| | | PARADIGM | 0·58 | 0·79 | 390 |
| | | TCGA | 0·61 | 0·83 | 356 |
| | | TRAIN II | 0·6 | 0·73 | 292 |

Table S4: Subset analysis showing Pearson's R and AUROC in separate patient subgroups. Each study shows a subset analysis for each clinical variable of interest, excluding missing data. Additionally, all studies are aggregated with a similarly subset analysis over all patients split by clinical variables of interest. The aggregate subset analysis by study matches the results presented in Figure 2. ER: Estrogen receptor; HER2: Human Epidermal growth factor Receptor 2, TN: Triple Negative; BC NST: Breast Cancer of No Special Type; ILC: Invasive Lobular Carcinoma.